\documentclass[preprint,12pt]{elsarticle}
\biboptions{sort&compress}

\usepackage{amsmath,amssymb}
\usepackage{graphicx}
\usepackage{epstopdf}
\usepackage{booktabs}
\usepackage{xcolor}
\usepackage[hidelinks]{hyperref}
\usepackage{lineno}

\newcommand{\dd}{\mathrm{d}}

\makeatletter\def\ps@pprintTitle{\let\@oddhead\@empty\let\@evenhead\@empty\def\@oddfoot{\footnotesize\itshape\hfill Preprint, \today}\let\@evenfoot\@oddfoot}\makeatother

\usepackage{longtable}
\begin{document}

\begin{frontmatter}

\title{Perilune margin and periapsis circulation of a maintenance-free low lunar orbit: tesseral forced eccentricity, insertion phase and the frozen center}

\author[tku]{Fu-Yuen Hsiao\corref{cor1}}
\ead{fyhsiao@mail.tku.edu.tw}
\author[tku]{Kaiti Wang}
\ead{ktwang@mail.tku.edu.tw}
\author[tku]{Wei-Cheng Chou}
\ead{tp.andychou@gmail.com}
\cortext[cor1]{Corresponding author}
\affiliation[tku]{organization={Department of Aerospace Engineering, Tamkang University},
            city={New Taipei City},
            postcode={251301},
            country={Taiwan}}

\begin{abstract}
A low eccentric lunar orbit ($60\times200$~km, $i=80.5^\circ$) left without maneuvers for three lunar years is studied for in situ sampling of the near-Moon plasma over a thick altitude shell. Its perilune margin is set by the monthly forced eccentricity of the odd-degree tesseral harmonics, the mechanism behind the translation theorem of low-lunar-orbit station keeping. We rederive it in Kaula form, give closed-form degree-3 amplitudes, and show that, in the tested single-averaged case, it lowers the minimum perilune by about 26~km, whereas the isolated $C_{22}$ effect is below 1~km. Because the insertion fixes the osculating and not the mean eccentricity, the lowest perilune depends on the node, from 31.6 to 58.8~km over twelve untargeted GMAT runs with a $50\times50$ field. A nonsingular single-averaged model reproduces the mission-minimum perilune altitude of all twenty GMAT runs within 0.5~km, and targeting the mean eccentricity raises the worst case to 39.9~km. The slow eccentricity vector rotates about a frozen center that converges only near zonal degree~50; the nominal eccentricity is close to twice the frozen center's distance from the origin and to the origin-crossing threshold of a reduced model (0.0374--0.0383), so the circulation is marginal: only 6 of the 20 runs belong to the circulating group, with marginal completion of one revolution, and an estimated 61--79\% of the shell cells are crossed. In the month-averaged model the frozen center approaches the origin closely near $76.3^\circ$ and $84.5^\circ$ inclination, where its signed coordinate changes sign, so the inclination, not only the eccentricity, is expected to control the circulation. The orbit plane precesses about an effective axis tilted by about $0.75^\circ$ from the spin axis; initial states and prediction models must nevertheless use a consistent lunar orientation model.
\end{abstract}


\end{frontmatter}

\section*{Nomenclature}
{\small\setlength{\LTpre}{2pt}\setlength{\LTpost}{6pt}
\begin{longtable}{@{}p{.27\linewidth}p{.70\linewidth}@{}}
\multicolumn{2}{@{}l}{\textit{Coordinate systems and angles}}\\[1pt]
$\mathcal L$, $\mathcal E$, $\mathcal M$ & coordinate systems: GMAT \texttt{luna} (lunar equator at J2000), lunar equator of date, DE405 principal axes (Moon-fixed); Section~\ref{sec:bg_mci} \\
$X_L,Y_L,Z_L$ & axes of $\mathcal L$ \\
$x_{\mathcal E},z_{\mathcal E}$; $x_{\mathcal M},z_{\mathcal M}$ & axes of $\mathcal E$ and $\mathcal M$ \\
$\mathsf C_1,\mathsf C_3$ & elementary rotation (direction-cosine) matrices about the first and third axes \\
$\Lambda$ & longitude of the node of the equator of date, measured from $X_L$, deg \\
$\beta$ & inclination of the equator of date to the J2000 lunar equator, deg \\
$\theta$ & rotation angle of the prime meridian, measured from $x_{\mathcal E}$, deg \\
$\hat{\mathbf z}$ & lunar spin axis of date \\
$\mathbf k$ & unit normal of the ecliptic plane (ecliptic pole) \\
$I$ & half-angle of the precession cone of $\hat{\mathbf z}$ about $\mathbf k$ ($1.54^\circ$) \\
$\gamma$ & precession rate of $\hat{\mathbf z}$, deg/day \\
$\mathbf w$ & effective precession axis, Eq.~\eqref{eq:bg_waxis} \\
$\hat{\mathbf h}$ & unit normal of the orbit plane (not to be confused with the altitude $h$) \\
\multicolumn{2}{@{}l}{\textit{Orbital elements and phases}}\\[1pt]
$a,e,i$ & semimajor axis (km), eccentricity, inclination (deg, from the equator of date) \\
$\Omega,\omega$ & longitude of the ascending node (from $x_{\mathcal E}$) and argument of perilune, deg \\
$M,f,u$ & mean anomaly, true anomaly, argument of latitude, deg \\
$n$ & mean motion, rad/s \\
$\varrho$ & semi-latus rectum $a(1-e^2)$, km \\
$r_p,r_a$ & perilune and apolune radii, km \\
$\psi$ & node phase $\Omega-\theta$, Eq.~\eqref{eq:bg_psi}, deg \\
$\nu_\psi$ & rate of $\psi$, deg/day \\
$\Omega_0,\omega_0,\theta_0,\psi_0$ & initial values of $\Omega,\omega,\theta,\psi$, deg \\
$i_0$, $\bar i$ & initial inclination (reported about $\mathbf w$) and mean inclination, deg \\
$\dot\Omega$, $\dot\omega_{\rm sec}$ & nodal rate of the orbit and apsidal rate of the slow motion, deg/day \\
$\Gamma_{J_2}$ & scale of the $J_2$ apsidal rate, deg/day \\
$\Delta\bar\omega$ & net rotation of the slow eccentricity vector over the mission, deg \\
$\mathcal G,\mathcal H$ & Delaunay momenta $\sqrt{\mu\varrho}$ and $\mathcal G\cos i$, km$^2$/s \\
\multicolumn{2}{@{}l}{\textit{Eccentricity vector}}\\[1pt]
$\varepsilon=\xi+\mathrm i\eta$ & nonsingular (complex) eccentricity vector $e\,\mathrm e^{\mathrm i\omega}$ \\
$\bar\varepsilon$, $\delta\varepsilon$ & slow (mean) part and monthly forced part of $\varepsilon$ \\
$\varepsilon_0$, $\bar\varepsilon_0$ & osculating and mean initial eccentricity vectors \\
$\varepsilon_c=\xi_c+\mathrm i\eta_c$ & frozen center of the slow motion; $e_c=|\varepsilon_c|$, $\eta_c$ its signed component along $\omega=90^\circ$ \\
$e_0$ & nominal initial eccentricity \\
$e_s$ & origin-crossing threshold of the reduced slow model \\
$\bar e_{\min}$ & smallest slow eccentricity over the mission \\
$M_e$ & eccentricity margin $\min_t|\bar\varepsilon|-\max_t|\delta\varepsilon|$ \\
\multicolumn{2}{@{}l}{\textit{Gravity field and perturbations}}\\[1pt]
$\mu$, $\mu_E$ & gravitational parameters of the Moon and the Earth, km$^3$/s$^2$ \\
$R_M$ & reference radius of the gravity field (1738.0~km) \\
$U$ & lunar gravitational potential, km$^2$/s$^2$ \\
$l,m$; $p,q$ & degree and order of a harmonic; indices of Kaula's expansion \\
$\bar C_{lm},\bar S_{lm}$, $\bar P_{lm}$ & fully normalized harmonic coefficients and associated Legendre functions \\
$J_l$, $J_{lm}$ & zonal coefficient and amplitude of the $(l,m)$ term (unnormalized) \\
$\bar R$, $Q_l$ & zonal disturbing function averaged over $M$, and its inclination--eccentricity integral, Eq.~\eqref{eq:thy_Rbar} \\
$R_{lmpq}$ & term of Kaula's expansion of the disturbing function \\
$F_{lmp},G_{lpq}$ & Kaula inclination and eccentricity functions \\
$\Theta_{lmpq}$ & argument of a term of Kaula's expansion \\
$\chi_{lm}$ & phase of the $(l,m)$ harmonic \\
$\kappa^\pm_{lm},\alpha^\pm_{lm}$ & coefficients of the forced eccentricity, Section~\ref{sec:forced} \\
$R_{\rm osc}$ & oscillating part of the disturbing function in $\psi$ \\
$R_E^{(2)}$ & quadrupole disturbing function of the Earth, Eq.~\eqref{eq:bg_quad} \\
$n_P$ & mean motion of the Earth about the Moon, rad/s \\
$\mathbf r_E$, $\lambda_E$ & selenocentric position and longitude of the Earth \\
\multicolumn{2}{@{}l}{\textit{Mission and design}}\\[1pt]
$\mathcal B$ & measurement region, Eq.~\eqref{eq:req_region} \\
$h,\varphi,\lambda$ & altitude above $R_{\rm alt}$ (km), latitude and longitude (deg) \\
$R_{\rm alt}$ & radius of the altitude reference sphere (1738.2~km) \\
$u_1,u_2$ & arguments of latitude at which a latitude $\varphi$ is crossed \\
$h_p$, $\bar h_p$ & perilune altitude and mean perilune altitude, km \\
$h_{\rm floor}$ & perilune floor (30~km) \\
$A$ & forced amplitude $a|\delta\varepsilon|_{\max}$, km \\
$\Delta h_{\rm slow}$ & decrease of the perilune altitude caused by the slow motion, km \\
$T$ & time horizon of the design quantity, days \\
R1--R5 & mission requirements (Table~\ref{tab:req_list}) \\
\end{longtable}}

\section{Introduction}
\label{sec:intro}

\paragraph{Motivation}
The Taiwan Space Agency (TASA) is developing its indigenous launch vehicle program and aims to make its first orbital attempt to place a 200~kg satellite into orbit in 2034 \cite{TaipeiTimes2025}. Because the launch infrastructure is still in the planning phase, accounting for potential lunar exploration in the 2040s provides a practical benchmark for sizing and scaling launch capacity, rather than designing solely around an initial small-satellite mission.

This paper constructs a notional lunar science mission in 2043 because the orbital geometry is particularly favorable. In that year, the inclination of the Moon's orbit relative to Earth's equator reaches the maximum of its 18.6-year cycle at approximately $28.6^\circ$ (the major lunar standstill) \cite{Meeus1998}. Because this alignment closely matches the $29.9^\circ$ parking orbit inclination associated with launches from the planned Taiwan launch site, the out-of-plane correction of the translunar trajectory is expected to be small; the translunar design of \cite{2025TASAmid2}, computed for this parking orbit, supports this, and the mission epoch is selected from its 2043 launch windows (Section~\ref{sec:req_epoch}). While this scenario fixes the baseline geometry for the case study, the orbital dynamics analyzed below are general and do not depend on this specific mission profile.

The science objective of the mission is the near-Moon plasma. Whistler-mode waves near 1~Hz have been reported around the Moon for decades \cite{Farrell1996,Nakagawa2003,Lou2023}. Lunar Prospector and Kaguya, both in polar orbits near 100~km altitude, observed narrowband and broadband whistlers whose power is concentrated above the crustal magnetic anomalies \cite{Halekas2006,Tsugawa2011,Tsugawa2012,Tsugawa2015}. The waves were suggested to be linked to regions of stronger magnetic fields, originating below the spacecraft and propagating upward. This assumption was adopted for the wave power transmission model described in Wang et al.\ \cite{2025Wang}, which solved the full wave equations for upward-propagating whistlers within a lunar ionosphere. This model is based on the crustal-field model of \cite{Ravat2020} and incorporates the photochemical density profiles of \cite{2016Choudhary}, along with electron and ion collisions, as outlined in the collisional treatment of \cite{WangChang2023}. They found that collisional attenuation can substantially reduce the transmitted wave power. The reduction depends on the wave frequency and on the plasma and magnetic-field conditions. A wave power spectrum acquired at 100~km is therefore not representative of the wave field near its source, and the attenuation itself conveys information on the plasma below the spacecraft.

For the mission considered here the same group extended the model to a three-dimensional map of the transmission ratio and the power spectral density between 30 and 200~km \cite{2026TASAmid4}. Scaled to the observed power spectral densities reported by the Kaguya mission at 100~km \cite{Tsugawa2011}, the largest spectral density at 30~km is preliminarily estimated to be about 12~nT$^2$/Hz above the strong crustal fields in the southern hemisphere. On the other hand, the electron density below 50~km also remains uncertain: ARTEMIS in-situ measurements limit it to about 35~cm$^{-3}$ near the equator \cite{2023Shen}, whereas radio occultation near the south pole implies values several times higher \cite{2016Choudhary}. These questions can only be answered by sampling the vertical structure of the near-Moon plasma, not a single altitude. The science requirement adopted for the orbit is to sample altitudes from 50 to 200~km over the mid-latitude belt $20^\circ\le|\varphi|\le70^\circ$, at all longitudes, during a single mission.

This requirement is the opposite of what low lunar orbits are usually designed for: the altitude at which a latitude is crossed must change during the mission. Section~\ref{sec:req} turns it into orbit specifications: a shell of 50--200~km altitude between $20^\circ$ and $70^\circ$ latitude to be sampled, a perilune that never falls below 30~km, an apolune below 250~km, and no orbit maintenance.

\paragraph{Literature review}
The gravity field of the Moon is strongly irregular, and for a low lunar orbiter the Earth acts as a third body whose tidal effect is far larger than that of the Moon on a low Earth satellite. Most low lunar orbits therefore change their eccentricity quickly and reach the surface within months unless they are maintained \cite{Konopliv2001,Lara2011}. Two remedies are common. A near-circular orbit is kept at its altitude by regular maneuvers. A frozen orbit is chosen so that the long-period variations of eccentricity and argument of periapsis vanish; such orbits have been studied in detail for the Moon \cite{ElyLieb2005,Ely2005,FoltaQuinn2006,Abad2009,NieGurfil2018} and for other bodies \cite{LiQiaoLi2019}. Both choices fix the altitude profile of the orbit. A circular orbit visits one altitude, and a frozen orbit always passes over a given latitude at the same altitude.

An eccentric orbit whose argument of periapsis $\omega$ changes monotonically moves its perilune and apolune through all latitudes, so each latitude is crossed at many altitudes during the mission. Circulating lunar orbits have been studied mostly at high altitude, where the Earth drives the apsidal motion \cite{RussellBrinckerhoff2009,McArdleRussell2021}. For a low orbit the circulation is driven by the lunar oblateness, and the question is whether the perilune stays above a safe floor over a full cycle.

The nonzonal field of the Moon is unusually large \cite{Konopliv2001}, and its effect on low orbits is well documented. High-resolution fields of degree 50 or more are recommended below 100~km \cite{SanJuan2020}. Lara \cite{Lara2025} showed that the coupling of the tesseral harmonics with the lunar rotation is amplified by the ratio $n/\dot\theta$, produces monthly patterns of the eccentricity vector, modifies the frozen-orbit equilibria, which in general no longer lie at the $\pm90^\circ$ of purely zonal models, and can turn an orbit that the mean dynamics predicts to be safe into an impacting one. In the operation of Lunar Reconnaissance Orbiter and in recent studies of extremely low lunar orbits, the eccentricity vector was found to follow a monthly pattern independent of its starting value---the ``translation theorem''---which Yarndley et al.\ \cite{Yarndley2025,Yarndley2026} traced to the odd-degree tesseral and sectoral harmonics and used for station keeping \cite{Holt2024}. These studies concern actively maintained low lunar orbits, near-circular and eccentric, including a mission case whose periapsis drifts naturally; the dependence of an unmaintained orbit on the insertion phase was not the subject of their test cases. Here we quantify the finite-mission coverage and perilune margin without routine maintenance and keep the insertion phase explicit. Bosanac and Miceli \cite{Bosanac2026} searched high-fidelity propagations of low lunar orbits for long-lifetime solutions with little drift of the perilune and compared them with analytical frozen-orbit predictions. The importance of the mean-to-osculating transformation for initializing low lunar orbits was emphasized by Nie and Gurfil \cite{NieGurfil2018}.

\paragraph{Research questions and approach}
What has not been addressed is how this mechanism constrains a maintenance-free orbit whose periapsis is meant to circulate through the whole latitude range. Four questions follow. (i) How far must the zonal series be carried to predict the long-term eccentricity of a $60\times200$~km orbit? (ii) What sets the perilune margin of such an orbit, and how does it depend on the insertion? (iii) How should the insertion be chosen? (iv) Does the mean periapsis actually circulate, and what limits the coverage? We answer them with a hierarchy of models: a doubly averaged zonal and Earth model for the design (Section~\ref{sec:theory}); a single-averaged model that keeps the lunar rotation and the tesseral field (Section~\ref{sec:savg}); a closed-form first-order solution for the forced eccentricity (Section~\ref{sec:forced}); and full-force GMAT propagations over three lunar years with a $50\times50$ field for verification (Section~\ref{sec:sim}). The single-averaged model rotates the field exactly as the propagator does, so that the two can be compared run by run.

\paragraph{Contributions}
\begin{enumerate}
\item A maintenance-free low eccentric orbit is analyzed for a shell-sampling science objective, with requirements on 330 latitude--altitude cells and explicit safety limits, and its perilune margin and coverage are quantified over three lunar years with twenty full-force propagations with a $50\times50$ field (Sections~\ref{sec:req}, \ref{sec:theory} and~\ref{sec:sim}).
\item The forced eccentricity of the odd-degree tesseral terms is rederived in Kaula form for this orbit, with closed-form amplitudes of the degree-3 terms that agree with numerical averaging at fixed elements within 2\% and with single-averaged propagation within 5\%. It is shown to be the dominant loss of perilune margin, while $C_{22}$, the Earth and the Sun are not (Sections~\ref{sec:savg}--\ref{sec:forced}).
\item The dependence of the lowest perilune on the insertion node is explained as the difference between osculating and mean initial eccentricity. A nonsingular single-averaged model reproduces the mission-minimum perilune altitude of all twenty runs within 0.5~km, and a two-step prediction without fitted parameters within 1.2~km for horizons up to 700 days, after which it overestimates the lowest perilune (Sections~\ref{sec:forced_node} and~\ref{sec:sim_scan}).
\item An insertion rule, targeting the mean eccentricity, is derived and tested in eight full-force runs, which quantify its effect on the node dependence and on the worst-case perilune (Sections~\ref{sec:forced_rule} and~\ref{sec:sim_target}).
\item The slow motion of the mean eccentricity vector is shown to be a rotation about a frozen center whose eccentricity converges only near degree~50 of the zonal series and is moved toward the origin by the Earth. For the nominal orbit the mean vector passes close to the origin, which makes the circulation of the mean periapsis and the coverage sensitive to the field truncation and to the node; we show that the nominal eccentricity lies close to the circular-contour estimate $2e_c$ and within the node-dependent range (0.0374--0.0383) of the origin-crossing threshold of a reduced slow model, and give a margin against near-circular passages, $\bar e_{\min}>|\delta\varepsilon|_{\max}$, separate from the check of the circulation itself (Section~\ref{sec:sim_circ}).
\item The results are reduced to three checks that need only the single-averaged model: the circulation of the slow phase, the margin against near-circular passages, and a perilune screen for targeted orbits that was below the observed lowest perilune in all eight targeted GMAT runs. The frozen center is shown to approach the origin closely at two inclinations, where its signed coordinate changes sign, which identifies the inclination as a second control of the circulation (Section~\ref{sec:design}).
\item We show that, for this orbit, the 18.6-year precession of the lunar spin axis makes the orbit plane precess about an effective axis inclined about $0.75^\circ$ to the spin axis, and that the inertial lunar frame of GMAT is fixed to the J2000 equator, $2.8^\circ$ from the 2043 equator; we show the spurious node dependence that these differences produce and how initial conditions and prediction models must be referred to the rotation model of the propagator (Sections~\ref{sec:bg_mci} and~\ref{sec:sim_frame}).
\end{enumerate}

\paragraph{Outline}
Section~\ref{sec:req} states the science requirements, the epoch and the candidate orbit classes. Section~\ref{sec:model} introduces the dynamical model and the reference frames, including the effective precession axis. Section~\ref{sec:theory} designs the orbit with the doubly averaged model. Section~\ref{sec:savg} develops and validates the single-averaged model and shows the convergence of the zonal series. Section~\ref{sec:forced} derives the forced eccentricity, explains the node dependence and gives the targeting rule. Section~\ref{sec:sim} verifies the design with GMAT. Section~\ref{sec:design} reduces the results to design checks. Section~\ref{sec:discussion} discusses the design implications and the limitations, and Section~\ref{sec:concl} concludes.

\section{Mission requirements and orbit selection}
\label{sec:req}

\subsection{Measurement region}
\label{sec:req_science}
The orbit must sweep the region
\begin{equation}
\mathcal B=\big\{(h,\varphi,\lambda):\;50\le h\le200~\mathrm{km},\;20^\circ\le|\varphi|\le70^\circ,\;0^\circ\le\lambda<360^\circ\big\},
\label{eq:req_region}
\end{equation}
where $h=r-R_{\rm alt}$ is the altitude above a sphere of radius $R_{\rm alt}=1738.2$~km, the lunar radius used by GMAT (the gravity-field reference radius $R_M=1738.0$~km of Section~\ref{sec:model} enters only the potential, and the nominal orbit is defined with it, Section~\ref{sec:theory}), $\varphi$ the selenographic latitude and $\lambda$ the longitude. The Moon rotates once in 27.3~days under an orbit plane that is nearly fixed over a month, which sweeps the longitude of the crossings, but this does not guarantee that every latitude--altitude combination is visited at every longitude. We design in the latitude--altitude plane; the longitude part of $\mathcal B$ is not verified in this paper. The coverage results of this paper refer to this marginal latitude--altitude coverage; a joint latitude--longitude--altitude requirement would need a separate recurrence analysis.

\subsection{Requirements}
\label{sec:req_orbit}
For design and verification $\mathcal B$ is divided into cells: the latitudes $\pm20^\circ$, $\pm25^\circ$, \dots, $\pm70^\circ$ (22 values) and altitude bins of 10~km between 50 and 200~km (15 bins), $22\times15=330$ cells. A cell is covered when the spacecraft crosses the latitude within the altitude bin at least once. One pass is enough, and short excursions outside the shell are allowed. The bin width is an analysis parameter and not a science requirement. Table~\ref{tab:req_list} lists the requirements. The perilune floor of 30~km is a hard safety limit. The apolune ceiling of 250~km is a preference, not a safety limit. Above the upper edge of the shell at 200~km the spacecraft collects no useful samples, and a higher apolune at the same perilune makes the orbit more eccentric, so that each revolution spans a wider range of altitudes over which the perturbing accelerations vary more, and the long-term evolution becomes harder to predict and maintain. R4 prescribes no mission duration: it asks only that $\omega$ sweep through one full circulation, so that every latitude is crossed at every altitude of the shell.

\begin{table}[htbp]
\centering
\caption{Orbit requirements.}
\label{tab:req_list}
\begin{tabular}{llp{8.6cm}}
\toprule
ID & Category & Requirement \\
\midrule
R1 & Coverage & Every one of the 330 latitude--altitude cells is crossed at least once. \\
R2 & Safety & Perilune altitude $\ge30$~km at all times. \\
R3 & Preference & Apolune altitude $\le250$~km at all times. \\
R4 & Circulation & At least one full circulation of $\omega$ during the mission. \\
R5 & Maintenance & No routine maneuvers. \\
\bottomrule
\end{tabular}
\end{table}

\subsection{Mission epoch}
\label{sec:req_epoch}
The epoch follows from the translunar design of the first phase of the project \cite{2025TASAmid2}. In that work the launch windows of 2043 from the Taiwan launch site were computed from the Earth--Moon geometry, the right ascension of the node attainable for a $29.91^\circ$ parking orbit, and the local sidereal time of the site. The arrival was placed where the Moon crosses the Earth's equatorial plane ($\omega_{\rm moon}+f_{\rm moon}=180^\circ$), so that the apoapsis of the last phasing loop meets the Moon in the equatorial plane. Among the thirteen monthly windows of 2043, the window of 2043-08-26 03:00~UTC was selected because the weather in Taiwan is favorable in August, the available launch time is only three hours from the ideal one (so the parking orbit is short), and it corresponds to 11:00 local time. A full-scenario simulation with phasing loops, the Earth's $J_2$ and trajectory correction reached the Moon on 2043-09-05 01:14~UTC, near new Moon, after 9.93~days and a total $\Delta V$ of 3.22~km/s. The science phase starts on 2043-09-10 00:00:00~UTC; the five days after arrival are reserved for the lunar orbit insertion and for in-orbit checkout of the spacecraft.
The propagations of Section~\ref{sec:sim} cover three lunar years of twelve synodic months each, $36\times29.53=1063$~days. The designed drift of $\omega$ is about $-0.35$ to $-0.4$~deg/day, so one circulation takes roughly 900--1000~days; the mission duration of three lunar years follows from this rate and is not a requirement in itself. Whether the circulation is completed depends on the slow motion near $e=0$ (Section~\ref{sec:sim_circ}).

\subsection{Candidate orbits}
\label{sec:req_candidates}
A circular orbit samples one altitude, so R1 needs a sequence of orbits joined by maneuvers, and R5 is violated. A frozen orbit has constant $(e,\omega)$; its altitude is a fixed function of latitude, $h=h(\varphi)$, and it traces a single curve in the $(\varphi,h)$ plane. It meets R2, R3 and R5 but covers only a thin strip of cells. In a drifting-periapsis orbit $\omega$ changes monotonically, and each latitude is visited at all altitudes between perilune and apolune. The price is a perilune altitude that varies along the circulation, so R2 has to be demonstrated. As Section~\ref{sec:sim_scan} shows, the full 330-cell coverage of R1 is also not reached within three lunar years, but the drifting orbit is the only class that covers most of the shell (Table~\ref{tab:req_compare}).

\begin{table}[htbp]
\centering
\small
\setlength{\tabcolsep}{2.5pt}
\caption{Qualitative comparison of the three orbit classes.}
\label{tab:req_compare}
\begin{tabular}{@{}lccc@{}}
\toprule
Requirement & Circular & Frozen & Drifting periapsis \\
\midrule
R1 coverage of $\mathcal B$ & No (one altitude) & No (one curve) & No (est.\ 61--79\%) \\
R2 perilune $\ge30$~km & Needs maneuvers & Yes & Tested cases$^a$ \\
R3 apolune $\le250$~km & Yes & Yes & Yes \\
R4 circulation of $\omega$ & -- & No & Not robust$^b$ \\
R5 no maneuvers & No & Yes & Yes \\
\bottomrule
\end{tabular}
\par\smallskip\footnotesize $^a$Met for the tested nominal initial conditions, with altitudes above the 1738.2~km reference sphere (lowest 31.6~km, Section~\ref{sec:sim}); not a terrain clearance or a guarantee under insertion errors. $^b$Node scan: 5 of 12 runs with the 27.3-day mean over days 14--1049, three of them within $5.7^\circ$ of one revolution, so that their classification changes with the filter (2--4 of 12 over a shorter common span); 6 of 20 runs including the targeted runs (Section~\ref{sec:sim_circ}).
\end{table}

\section{Dynamical model}
\label{sec:model}

\subsection{Frames, coordinate systems and node phase}
\label{sec:bg_mci}
\paragraph{Frames}
Two reference frames are used, both centered at the Moon: an inertial (non-rotating) frame and a body-fixed frame that rotates with the Moon.

\begin{figure}[htbp]
\centering
\includegraphics[width=0.85\textwidth]{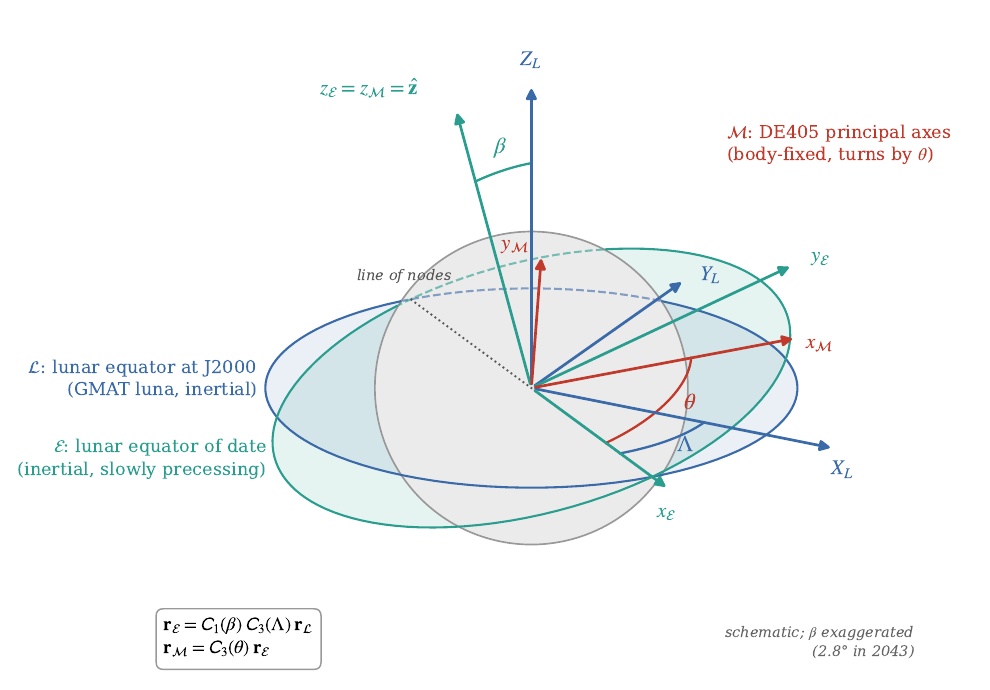}
\caption{Coordinate systems $\mathcal L$, $\mathcal E$ and $\mathcal M$ of Table~\ref{tab:frames} (schematic; the gray sphere is the Moon, and the tilt $\beta$ is exaggerated, $2.8^\circ$ in 2043). $\mathcal E$ follows from $\mathcal L$ by a rotation $\Lambda$ about $Z_L$, which brings $X_L$ to the node $x_{\mathcal E}$ of the equator of date, and a rotation $\beta$ about $x_{\mathcal E}$; $\mathcal M$ follows from $\mathcal E$ by the rotation $\theta$ of the prime meridian about $z_{\mathcal E}=\hat{\mathbf z}$, Eq.~\eqref{eq:bg_rot}.}
\label{fig:frames}
\end{figure}

\paragraph{Coordinate systems}
Three coordinate systems are defined in these frames; they are listed in Table~\ref{tab:frames} and sketched in Fig.~\ref{fig:frames}.
\begin{itemize}
\item $\mathcal L$ (inertial frame), the GMAT \texttt{luna} system (\texttt{BodyInertial} axes): the $X_LY_L$ plane is the lunar equator at J2000 and $Z_L$ the lunar pole at J2000. Its axes are fixed.
\item $\mathcal E$ (inertial frame, slowly precessing axes), the lunar equator of date: $z_{\mathcal E}=\hat{\mathbf z}$, the lunar spin axis of date; $x_{\mathcal E}$ points to the ascending node of the equator of date on the $X_LY_L$ plane; $y_{\mathcal E}=z_{\mathcal E}\times x_{\mathcal E}$. These axes do not rotate with the Moon; they follow only the slow precession of $\hat{\mathbf z}$ described below, about $1.5^\circ$ over the three-year mission. The precession is negligible for a mission of one or two months, but it must be taken into account over three years.
\item $\mathcal M$ (body-fixed frame), the DE405 principal-axis system: $z_{\mathcal M}$ is the spin axis of the DE405 libration model and $x_{\mathcal M}$ the principal axis that defines the prime meridian. With the monthly physical libration of $z_{\mathcal M}$ ($\lesssim0.03^\circ$) averaged out, $z_{\mathcal M}=\hat{\mathbf z}$.
\end{itemize}

\paragraph{Rotations}
With the elementary rotation (direction-cosine) matrices $\mathsf C_1$ and $\mathsf C_3$ about the first and third axes, the coordinates of a vector in the three systems are related by
\begin{equation}
\mathbf r_{\mathcal E}=\mathsf C_{\mathcal{EL}}(t)\,\mathbf r_{\mathcal L},\quad \mathsf C_{\mathcal{EL}}(t)=\mathsf C_1\big(\beta(t)\big)\,\mathsf C_3\big(\Lambda(t)\big),\qquad
\mathbf r_{\mathcal M}=\mathsf C_3\big(\theta(t)\big)\,\mathbf r_{\mathcal E},
\label{eq:bg_rot}
\end{equation}
where $\Lambda$ is the longitude of the node of the equator of date measured from $X_L$, $\beta$ the inclination of the equator of date to the J2000 lunar equator, and $\theta$ the rotation angle of the prime meridian measured from $x_{\mathcal E}$. The angles $\Lambda$ and $\beta$ change slowly with the precession of $\hat{\mathbf z}$ ($\beta=2.8^\circ$ in 2043); $\theta$ advances by one turn per sidereal month.

\paragraph{Computations and systems}
The GMAT propagations, their initial Cartesian states (\ref{app:states}) and their output are expressed in $\mathcal L$. The single-averaged equations of Section~\ref{sec:savg} are integrated for elements $a,e,i,\Omega,\omega$ referred to $\mathcal L$; at each evaluation of the rates the position is rotated into $\mathcal M$, where the gravity field (LP165P) is evaluated, and the acceleration is rotated back, $\mathbf a_{\mathcal L}=\mathsf C_{\mathcal{ML}}^{\mathsf T}\,\mathbf a_{\mathcal M}(\mathsf C_{\mathcal{ML}}\mathbf r_{\mathcal L})$ with $\mathsf C_{\mathcal{ML}}=\mathsf C_3(\theta)\,\mathsf C_{\mathcal{EL}}$. The analytical theory of Sections~\ref{sec:model}--\ref{sec:forced} refers the elements to $\mathcal E$: $i$ is measured from the equator of date, as the zonal and Kaula expansions and the latitude formulas require, and $\Omega$ and the longitude $\lambda_E$ of the Earth are measured from $x_{\mathcal E}$. Latitudes, coverage and the slow eccentricity vector of Table~\ref{tab:sim_scan} are also referred to $\mathcal E$.

\paragraph{Effective precession axis}
The spin axis $\hat{\mathbf z}$ precesses about the ecliptic pole, the unit normal $\mathbf k$ of the ecliptic plane, with the 18.6-year period of the lunar node, on a cone of half-angle $I=1.54^\circ$ \cite{Archinal2018}; over the three-year mission it moves along a $56^\circ$ arc of this cone, that is, by about $1.5^\circ$. The precession rate, $\gamma=-0.053$~deg/day, is only three times smaller than the nodal rate of the spacecraft orbit, that is, the regression of its ascending node on the equator of date ($\dot\Omega\approx-0.16$~deg/day; it is driven mainly by $J_2$, which alone gives $-0.19$~deg/day, Section~\ref{sec:thy_zonal}), so the orbit plane does not follow the moving equator adiabatically. If the precession of $\hat{\mathbf z}$ is taken as uniform and the nodal rate as constant, the orbit normal $\hat{\mathbf h}$ obeys $\dot{\hat{\mathbf h}}=\dot\Omega\,\hat{\mathbf z}\times\hat{\mathbf h}$ in the inertial frame and, in axes turning with angular velocity $\gamma\,\mathbf k$ in which $\hat{\mathbf z}$ is fixed, $\dot{\hat{\mathbf h}}=(\dot\Omega\,\hat{\mathbf z}-\gamma\,\mathbf k)\times\hat{\mathbf h}$. In this approximation $\hat{\mathbf h}$ precesses uniformly about the axis
\begin{equation}
\mathbf w\propto\dot\Omega\,\hat{\mathbf z}-\gamma\,\mathbf k ,\qquad \mathbf w\cdot\hat{\mathbf z}>0,
\label{eq:bg_waxis}
\end{equation}
which is fixed in these co-precessing axes and inclined to $\hat{\mathbf z}$ by $\approx \gamma\sin I/(\dot\Omega-\gamma\cos I)\approx0.75^\circ$ for the nominal orbit (0.74--0.88$^\circ$ for the nodal rates $-0.164$ to $-0.146$~deg/day of the GMAT runs), on the side away from $\mathbf k$. The angle between $\hat{\mathbf h}$ and $\mathbf w$ is then constant, whereas the inclination to the equator of date oscillates by up to $\pm0.75^\circ$ over a nodal cycle. We call $\mathbf w$ the effective precession axis. It is an approximate construction for a given orbit, not an exact integral of the full problem; in the zonal-only run D1, however, $\sqrt{1-e^2}\cos i$ measured from $\mathbf w$ stays constant to $3\times10^{-5}$ over three lunar years (Section~\ref{sec:sim_frame}). The axis $\mathbf w$ is used only to describe the motion of the orbit plane and to report the initial inclinations of the GMAT runs.

\paragraph{Reference axes for the initial inclination}
The GMAT runs, which are defined in Section~\ref{sec:sim} (Table~\ref{tab:sim_runs}), start from states whose initial inclination is measured from one of three axes close to $\hat{\mathbf z}$: an axis fitted to the zonal-only run D1, the pole of the IAU rotation model at 2043-09-10, and the effective precession axis $\mathbf w$ of Eq.~\eqref{eq:bg_waxis}. The states are built about the first two, as listed in Table~\ref{tab:frames}, and the initial inclinations $i_0$ of all runs are reported about $\mathbf w$; Section~\ref{sec:sim_frame} explains the choice.

\begin{table}[htbp]
\centering
\small
\caption{Coordinate systems and reference axes used in the paper. All are centered at the Moon. The GMAT run groups are defined in Table~\ref{tab:sim_runs}. Epochs are given in UTC; GMAT evaluates the ephemerides and the lunar orientation in TDB.}
\label{tab:frames}
\begin{tabular}{@{}p{0.08\textwidth}p{0.11\textwidth}p{0.32\textwidth}p{0.37\textwidth}@{}}
\toprule
System & Frame & Axes & Used for \\
\midrule
$\mathcal L$ & inertial & GMAT \texttt{luna} (\texttt{BodyInertial}): $X_LY_L$ plane = lunar equator at J2000, $Z_L$ = J2000 pole & GMAT propagation, initial states and output; elements of the single-averaged model \\
$\mathcal E$ & inertial, slowly precessing & $z_{\mathcal E}=\hat{\mathbf z}$ (spin axis of date); $x_{\mathcal E}$ = node of the equator of date on $X_LY_L$ & $i,\Omega,\omega,\psi,\lambda_E$ in Sections~\ref{sec:model}--\ref{sec:forced}; latitudes, coverage; slow eccentricity vector \\
$\mathcal M$ & body-fixed & DE405 principal axes; $z_{\mathcal M}=\hat{\mathbf z}$, $x_{\mathcal M}$ at angle $\theta$ from $x_{\mathcal E}$ & evaluation of the gravity field (LP165P) \\
\midrule
\multicolumn{2}{@{}l}{Reference axis} & Definition & Used for \\
\midrule
\multicolumn{2}{@{}l}{$\mathbf w$} & effective precession axis, Eq.~\eqref{eq:bg_waxis} & reported $i_0$ of the GMAT runs; orbit-plane motion \\
\multicolumn{2}{@{}l}{fitted axis} & linear fit to run D1 & initial states of runs T, E, F, G, M, H \\
\multicolumn{2}{@{}l}{IAU pole} & IAU rotation model at 2043-09-10 & initial states of runs D, I, S \\
\bottomrule
\end{tabular}
\end{table}

\paragraph{Node phase}
Because the Moon is tidally locked, its rotation angle $\theta$ follows the longitude $\lambda_E$ of the Earth up to librations. The tesseral potential depends on the node through $\Omega-\theta$, and the Earth tide through $\Omega-\lambda_E$. We use the phase
\begin{equation}
\psi=\Omega-\theta ,
\label{eq:bg_psi}
\end{equation}
which completes one cycle in a sidereal month while $\Omega$ changes slowly.

\subsection{Gravity field}
\label{sec:bg_gravity}
\begin{figure}[htbp]
\centering
\includegraphics[width=0.98\textwidth]{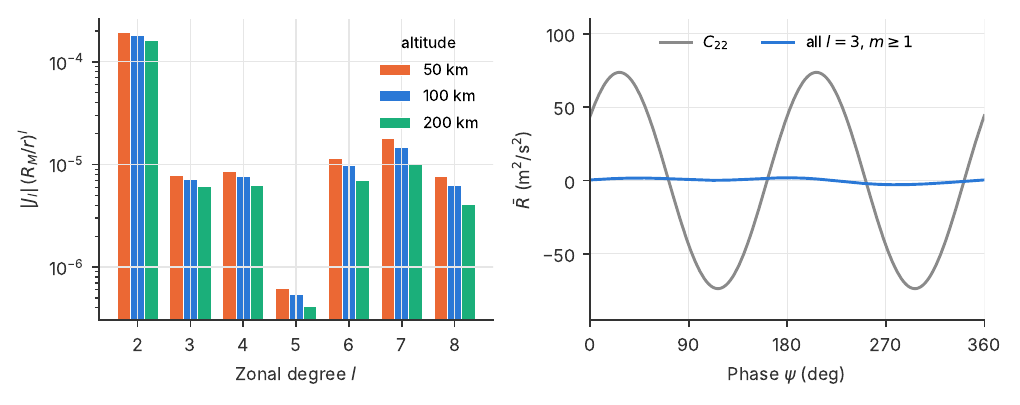}
\caption{Relative size of the low-degree gravity terms. Left: $|J_l|(R_M/r)^l$ for $l=2$--8 at three altitudes. Right: singly averaged potential at $a=1868$~km, $e=0.0375$, $i=80.5^\circ$ versus $\psi$ for $C_{22}$ and for all $l=3$ terms.}
\label{fig:bg_gravity_terms}
\end{figure}

The lunar potential is expanded in fully normalized spherical harmonics,
\begin{equation}
U=\frac{\mu}{r}\left[1+\sum_{l=2}^{\infty}\Big(\frac{R_M}{r}\Big)^{l}\sum_{m=0}^{l}\bar P_{lm}(\sin\varphi)\big(\bar C_{lm}\cos m\lambda+\bar S_{lm}\sin m\lambda\big)\right],
\label{eq:bg_potential}
\end{equation}
with the LP165P coefficients \cite{Konopliv2001}, $\mu=4902.801056$~km$^3$/s$^2$ and reference radius $R_M=1738.0$~km. We write $J_l=-\sqrt{2l+1}\,\bar C_{l0}$ and use the convention ``force $=\nabla R$''. In orbital elements Kaula's expansion \cite{Kaula1966} gives terms with arguments
\begin{equation}
\Theta_{lmpq}=(l-2p)\omega+(l-2p+q)M+m(\Omega-\theta).
\label{eq:bg_Theta}
\end{equation}
Averaging over the mean anomaly keeps the terms with $l-2p+q=0$: the zonal terms ($m=0$), with argument $(l-2p)\omega$, and the tesseral terms, with argument $(l-2p)\omega+m\psi$. For the orbit of this paper $n/\dot\theta\approx326$, so the tesseral terms are nonresonant. Table~\ref{tab:bg_coeff} lists the low-degree coefficients, and Fig.~\ref{fig:bg_gravity_terms} compares their contributions at the orbit.

\begin{table}[htbp]
\centering
\small
\caption{Unnormalized LP165P coefficients of low degree.}
\label{tab:bg_coeff}
\begin{tabular}{@{}lcp{5.0cm}@{}}
\toprule
Coefficient & Value & Role for the science orbit \\
\midrule
$J_2$ & $2.0324\times10^{-4}$ & drives the circulation of $\omega$ and $\Omega$ \\
$J_3$, $J_4$ & $8.476\times10^{-6}$, $-9.592\times10^{-6}$ & shape of the $(e,\omega)$ contour \\
$C_{22}$, $S_{22}$ & $2.2357\times10^{-5}$, $1.08\times10^{-8}$ & monthly oscillation of $i$ \\
$C_{31}$, $S_{31}$ & $2.8437\times10^{-5}$, $5.902\times10^{-6}$ & forced eccentricity (largest single pair) \\
$C_{32}$, $S_{32}$ & $4.846\times10^{-6}$, $1.671\times10^{-6}$ & forced eccentricity \\
$C_{33}$, $S_{33}$ & $1.713\times10^{-6}$, $-2.49\times10^{-7}$ & forced eccentricity \\
$C_{21}$, $S_{21}$ & $\sim10^{-9}$ & negligible \\
\bottomrule
\end{tabular}
\end{table}

\subsection{Third bodies and full force model}
\label{sec:bg_3body}
For the Earth at selenocentric position $\mathbf r_E$, the quadrupole part of the disturbing function is
\begin{equation}
R_E^{(2)}=\frac{n_P^2}{2}\Big[3(\mathbf r\cdot\hat{\mathbf u})^2-r^2\Big],\qquad n_P^2=\frac{\mu_E}{r_E^3},
\label{eq:bg_quad}
\end{equation}
with $n_P=2.6653\times10^{-6}$~rad/s. The solar tide is 0.56\% of the Earth's. The numerical propagations integrate
\begin{equation}
\ddot{\mathbf r}=-\frac{\mu}{r^3}\mathbf r+\nabla R_{grav}+\sum_{P\in\{E,S\}}\mu_P\left[\frac{\mathbf r_P-\mathbf r}{|\mathbf r_P-\mathbf r|^3}-\frac{\mathbf r_P}{r_P^3}\right]+\mathbf a_{SRP}+\mathbf a_{rel},
\label{eq:bg_eom}
\end{equation}
with the settings of Table~\ref{tab:bg_forces}.

\begin{table}[htbp]
\centering
\small
\caption{Force model and integrator of the GMAT propagations.}
\label{tab:bg_forces}
\begin{tabular}{@{}p{0.27\textwidth}p{0.62\textwidth}@{}}
\toprule
Item & Setting \\
\midrule
Lunar gravity & LP165P, degree and order 50 for the main runs (T, M, H), 20 for the decomposition and term-group runs and for a repeat of the main runs, no tide \\
Third bodies & Earth and Sun, point masses, JPL DE405 ephemeris \cite{Standish1998} \\
Solar radiation pressure & spherical, $C_r=1.8$, area 1~m$^2$, mass 850~kg ($A/m=1.2\times10^{-3}$~m$^2$/kg) \\
Relativistic correction & on \\
Integrator & RKF89, accuracy $10^{-11}$, step 60--2700~s \\
Output & osculating state at every periapsis and apoapsis passage (about 12\,700 orbits) \\
\bottomrule
\end{tabular}
\end{table}

\section{Doubly averaged design model}
\label{sec:theory}

The orbit has three separated time scales: the orbital period (2.01~h), the lunar month (27.3~days), and the circulation of $\omega$ and $\Omega$ (years). The classical design model averages over both fast angles. We state it first because it gives the circulation rate and the structure of the coverage; Sections~\ref{sec:savg} and~\ref{sec:forced} then show what it misses.

\subsection{Zonal potential}
\label{sec:thy_zonal}
With the zonal harmonics up to degree $N$, the potential averaged over the mean anomaly is
\begin{gather}
\bar R=-\frac{\mu}{a}\sum_{l=2}^{N}J_l\Big(\frac{R_M}{a}\Big)^{l}(1-e^2)^{\frac12-l}Q_l,\nonumber\\
Q_l=\frac{1}{2\pi}\int_0^{2\pi}(1+e\cos f)^{l-1}P_l\big(\sin i\,\sin(\omega+f)\big)\,\dd f .
\label{eq:thy_Rbar}
\end{gather}
The quadrature form agrees with the Legendre-addition series of \cite{LiQiaoLi2019} to $10^{-15}$ and needs no truncation in $e$. From Lagrange's planetary equations the averaged rates are
\begin{equation}
\dot\omega=\frac{\sqrt{1-e^2}}{na^2e}\,\frac{\partial\bar R}{\partial e}-\frac{\cot i}{na^2\sqrt{1-e^2}}\,\frac{\partial\bar R}{\partial i},\qquad
\dot e=-\frac{\sqrt{1-e^2}}{na^2e}\,\frac{\partial\bar R}{\partial\omega}.
\label{eq:thy_rates}
\end{equation}
For $J_2$ alone,
\begin{equation}
\dot\omega_{J_2}=\tfrac34\,nJ_2\Big(\frac{R_M}{\varrho}\Big)^{2}\big(5\cos^2i-1\big),\qquad
\dot\Omega_{J_2}=-\tfrac32\,nJ_2\Big(\frac{R_M}{\varrho}\Big)^{2}\cos i ,
\label{eq:bg_secularJ2}
\end{equation}
with the semi-latus rectum $\varrho=a(1-e^2)$; the drift is retrograde for $i>63.4^\circ$, and at the nominal orbit $\dot\omega_{J_2}=-0.49$~deg/day and $\dot\Omega_{J_2}=-0.19$~deg/day. With the full zonal series and the Earth the mean nodal rate is about $-0.16$~deg/day ($-0.146$ to $-0.164$~deg/day in the GMAT runs), the value used in Sections~\ref{sec:bg_mci} and~\ref{sec:sim}.

\subsection{Earth tide}
\label{sec:thy_earth}
Averaging Eq.~\eqref{eq:bg_quad} over the orbit and over the motion of the Earth on a circular orbit in the lunar equatorial plane gives \cite{McArdleRussell2021}
\begin{equation}
\langle\langle R_E^{(2)}\rangle\rangle=\frac{n_P^2a^2}{16}\Big[(2+3e^2)(3\cos^2i-1)+15e^2\sin^2i\cos2\omega\Big],
\label{eq:thy_Earth}
\end{equation}
which adds about $-0.09$~deg/day to $\dot\omega$ at $\omega=90^\circ$ and modulates the rate along the circulation.

\subsection{Energy contours, drift-rate bound and coverage}
\label{sec:thy_contour}
Without $\Omega$ and $M$ in the doubly averaged potential, $\sqrt{1-e^2}\cos i$ and $R_{tot}=\bar R+\langle\langle R_E^{(2)}\rangle\rangle$ are constants, so the orbit moves on a closed curve in the $(e,\omega)$ plane, and the circulation period is $T_\omega=\oint\dd\omega/\dot\omega$. Since $5\cos^2i-1\in[-1,4]$, the $J_2$ drift rate is bounded,
\begin{equation}
-\Gamma_{J_2}\le\dot\omega_{J_2}\le4\Gamma_{J_2},\qquad \Gamma_{J_2}=\tfrac34 nJ_2\Big(\frac{R_M}{\varrho}\Big)^2=0.57~\mathrm{deg/day}
\label{eq:thy_bound}
\end{equation}
at $a=1868$~km, so, from the $J_2$ contribution alone, a retrograde drift faster than about $-0.6$~deg/day is not available at this semi-major axis.

A latitude $\varphi$ is crossed at $u_1=\arcsin(\sin\varphi/\sin i)$ and $u_2=\pi-u_1$, at the altitude
\begin{equation}
h(\varphi,\omega)=\frac{a(1-e^2)}{1+e\cos(u-\omega)}-R_{\rm alt} .
\label{eq:thy_h}
\end{equation}
On a circulating orbit one branch covers the shell at latitude $\varphi$ if $\min_\omega h\le h_{lo}$ and $\max_\omega h\ge h_{hi}$; for fixed elements the minimum occurs when perilune is over the latitude and the maximum when apolune is. When the orbital elements vary, the crossing-altitude extrema need not occur at apsidal passage. In the design model, coverage of a cell therefore depends mainly on the value of $e$ at the time the apses pass the latitude, i.e.\ on the shape of the contour; the full-force counts of Section~\ref{sec:sim_scan} evaluate every revolution instead.

\subsection{Monthly inclination oscillation}
\label{sec:thy_raan}
The singly averaged Earth quadrupole and the sectorial term
\begin{equation}
\langle R_{22}\rangle=\frac{3\mu R_M^2C_{22}\sin^2 i}{2a^3(1-e^2)^{3/2}}\cos2\psi
\label{eq:bg_C22avg}
\end{equation}
oscillate with $2\psi$. Writing their sum as $R_{\rm osc}(\psi)$ and using $\dot{\mathcal H}=\partial R_{\rm osc}/\partial\psi$ for the Delaunay momentum $\mathcal H=\mathcal G\cos i$, $\mathcal G=\sqrt{\mu a(1-e^2)}$, with $\dot\psi=\nu_\psi\approx-\dot\theta$, first-order averaging gives
\begin{equation}
\delta i\simeq-\frac{1}{\sin i}\,\frac{R_{\rm osc}(\psi)}{\nu_\psi\,\mathcal G},\qquad
\bar i\approx i_0-0.606^\circ\cos2\psi_0 ,
\label{eq:thy_dI}
\end{equation}
of which $C_{22}$ accounts for $0.54^\circ$ and the Earth for $0.07^\circ$ \cite{Lara2025}. The full-force runs of Section~\ref{sec:sim} show this oscillation: the difference between their mean and initial inclinations contains a $2\Omega_0$ harmonic of amplitude $0.48^\circ$ with the phase of Eq.~\eqref{eq:thy_dI}, together with a smaller $\Omega_0$ harmonic of $0.24^\circ$. Because $C_{22}$ does not depend on $\omega$, it does not change $e$ at first order; Section~\ref{sec:forced} shows that it is not the term that limits the perilune.

\subsection{Nominal design}
\label{sec:req_nominal}
The perilune altitude is set to 60~km so that 30~km remain above the floor for insertion and navigation errors; the apolune is 200~km. With $r_p=1798$~km and $r_a=1938$~km, $a=1868$~km and $e=0.0374732334$. The inclination must exceed $70^\circ$ to reach the highest latitude and $63.4^\circ$ for a retrograde drift. We take $i=80.5^\circ$, $\omega_0=90^\circ$ and $f_0=0$ (Table~\ref{tab:req_nominal}). The initial node is set by the transfer.

\begin{table}[htbp]
\centering
\small
\caption{Nominal orbit (osculating elements at the epoch, inclination with respect to the lunar equator of the epoch).}
\label{tab:req_nominal}
\begin{tabular}{lc}
\toprule
Element or quantity & Value \\
\midrule
Perilune $\times$ apolune altitude & $60\times200$~km \\
Semi-major axis $a$, eccentricity $e$ & 1868~km, 0.0374732334 \\
Inclination $i$ & $80.5^\circ$ \\
$\omega_0$, $f_0$ & $90^\circ$, $0^\circ$ \\
Initial node $\Omega_0$ & set by the transfer \\
Orbital period & 2.01~h \\
Drift of $\omega$ (GMAT, where it circulates) & $-0.34$ to $-0.35$~deg/day \\
Epoch & 2043-09-10 00:00:00 UTC \\
\bottomrule
\end{tabular}
\end{table}

The inclination $80.5^\circ$ was originally selected with a doubly averaged model truncated at $J_8$, in which the full 330-cell coverage was obtained between $80.25^\circ$ and $81.0^\circ$. Table~\ref{tab:thy_window} repeats the computation with the converged zonal series (Section~\ref{sec:savg_zonal}) and the Earth.

\begin{table}[htbp]
\centering
\small
\caption{Zonal series $J_2$--$J_{30}$ (converged for the lowest perilune; the center of the slow motion needs degree~50, Section~\ref{sec:sim_circ}) plus Earth, single-averaged over the mean anomaly without tesseral terms: lowest and highest altitude, cells covered (mean elements) and total change of $\omega$ over three lunar years, versus the initial inclination ($a=1868$~km, $e_0=0.0375$, $\omega_0=90^\circ$, $\Omega_0=90^\circ$).}
\label{tab:thy_window}
\begin{tabular}{lcccc}
\toprule
$i_0$ (deg) & Lowest (km) & Highest (km) & Cells (of 330) & $\Delta\omega$ (deg) \\
\midrule
79.25 & 57.8 & 201.8 & 246 & $-393$ \\
79.50 & 57.8 & 201.8 & 235 & $-394$ \\
79.75 & 57.9 & 201.7 & 229 & $-396$ \\
80.00 & 58.0 & 201.6 & 224 & $-397$ \\
80.25 & 58.3 & 201.3 & 218 & $-398$ \\
80.50 & 58.6 & 201.0 & 211 & $-399$ \\
80.75 & 59.0 & 200.6 & 208 & $-400$ \\
81.00 & 59.5 & 200.1 & 205 & $-402$ \\
81.25 & 59.8 & 199.8 & 205 & $-405$ \\
81.50 & 59.8 & 199.8 & 204 & $-408$ \\
82.00 & 59.8 & 199.8 & 206 & $-417$ \\
\bottomrule
\end{tabular}
\end{table}

Two conclusions differ from the truncated model. With the converged series the lowest perilune of the zonal problem stays within 2~km of the initial 59.8~km for every inclination between $79.25^\circ$ and $82^\circ$, so the zonal dynamics does not limit the margin. On the other hand the mean elements cover only 62--75\% of the cells (with $J_2$--$J_{30}$; the center of the slow motion needs degree~50, Section~\ref{sec:sim_circ}, so these values and the trends with inclination below are indicative only), because $e$ does not vary enough along the contour to bring the apses to the edges of the shell at every latitude; lower inclinations cover slightly more. The inclination therefore controls the drift rate and the coverage, not the safety margin, which is set by the nonzonal terms (Section~\ref{sec:forced}). Here and in the rest of the paper, altitudes, cells, perilune bounds and the safety requirement are measured above $R_{\rm alt}=1738.2$~km, the lunar radius used by GMAT, so that the nominal $60\times200$~km orbit (defined with $R_M=1738.0$~km) starts at $59.8\times199.8$~km.

\section{Averaged dynamics with the lunar rotation retained}
\label{sec:savg}

\subsection{Single averaging}
\label{sec:savg_model}
The orbital period ($2.01$~h) is short compared with the rotation period of the Moon ($27.32$~days), whereas the rotation is fast compared with the circulation of $\omega$ and $\Omega$ (years). We therefore average only over the mean anomaly and keep the rotation angle $\theta(t)$ of the Moon explicit. The body axes are those of the propagator used for verification: the principal-axis frame given by the DE405 libration angles \cite{Standish1998}, in which the LP165P coefficients are defined. The model was implemented in Python by the authors with the assistance of a generative AI tool (Claude, Anthropic); the results of the model for the GMAT test cases are compared with the independent propagations of Section~\ref{sec:sim}. For a disturbing acceleration $\mathbf f$ with radial, transverse and normal components $(f_R,f_T,f_N)$, the averaged rates are
\begin{equation}
\dot{\bar x}=\frac{1}{2\pi}\int_0^{2\pi}\dot x\big(a,e,i,\Omega,\omega,M;\theta(t)\big)\,\dd M ,
\label{eq:savg_avg}
\end{equation}
where $\dot x$ is any of the Gauss equations,
\begin{align}
\dot e&=\frac{1}{\mathcal G}\Big[\varrho\sin f\,f_R+\big((\varrho+r)\cos f+re\big)f_T\Big],\nonumber\\
\dot\imath&=\frac{r\cos u}{\mathcal G}f_N,\nonumber\\
\dot\Omega&=\frac{r\sin u}{\mathcal G\sin i}f_N,\label{eq:savg_gauss}\\
\dot\omega&=\frac{1}{\mathcal G e}\Big[{-\varrho}\cos f\,f_R+(\varrho+r)\sin f\,f_T\Big]-\frac{r\sin u\cos i}{\mathcal G\sin i}f_N,\nonumber
\end{align}
with $\varrho=a(1-e^2)$, $\mathcal G=\sqrt{\mu\varrho}$ and $u=\omega+f$. The integral~\eqref{eq:savg_avg} is evaluated by quadrature at fixed $\theta$, with the full spherical-harmonic field of LP165P \cite{Konopliv2001} in the body frame and, when required, the Earth as a point mass at its ephemeris position. The resulting equations contain the slow angles $\omega,\Omega$ and the monthly angle
\begin{equation}
\psi=\Omega-\theta ,\qquad \dot\psi=\nu_\psi\simeq\dot\Omega-\dot\theta\approx-0.233~\mathrm{rad/day},
\label{eq:savg_psi}
\end{equation}
and are integrated with a fixed-step fourth-order Runge--Kutta scheme with a step of 0.5~day; halving the step changes the lowest altitude of T060 and T090 ($50\times50$) by at most 0.13~km and leaves the cell count, the smallest slow eccentricity and the circulation classification unchanged. Because the slow eccentricity of the orbits studied here passes close to zero, the equations are integrated in the nonsingular eccentricity vector $\varepsilon=\xi+\mathrm{i}\eta=e\,\mathrm{e}^{\mathrm{i}\omega}$ instead of $(e,\omega)$: $\dot\varepsilon=(\dot e+\mathrm{i}\,e\dot\omega)\,\mathrm{e}^{\mathrm{i}\omega}$, in which $e\dot\omega$ from Eq.~\eqref{eq:savg_gauss} is regular at $e=0$; the angle $\omega=\arg\varepsilon$ is used only to place the quadrature points along the orbit, whose geometry becomes independent of $\omega$ as $e\to0$, so the rates remain continuous. The quadrature uses 144 points in $M$, which resolves the field to degree~50; with 72 points the rates change by less than $10^{-9}$ of their size. In terms of Kaula's expansion \cite{Kaula1966}, Eq.~\eqref{eq:savg_avg} keeps every term with argument
\begin{equation}
\Theta_{lmp}=(l-2p)\,\omega+m\,\psi ,
\label{eq:savg_Theta}
\end{equation}
i.e.\ all zonal terms ($m=0$) and all tesseral terms with $l-2p+q=0$, which vary with the lunar rotation. The doubly averaged model of Section~\ref{sec:theory} is recovered by dropping all terms with $m\neq0$ and, in addition, averaging the Earth term over the motion of the Earth, taken on a circular orbit in the lunar equatorial plane, as in Eq.~\eqref{eq:thy_Earth}; with the Earth on its ephemeris orbit, the single-averaged model keeps the monthly Earth terms.

\subsection{Validation against full-force propagation}
\label{sec:savg_valid}
\begin{figure}[htbp]
\centering
\includegraphics[width=0.98\textwidth]{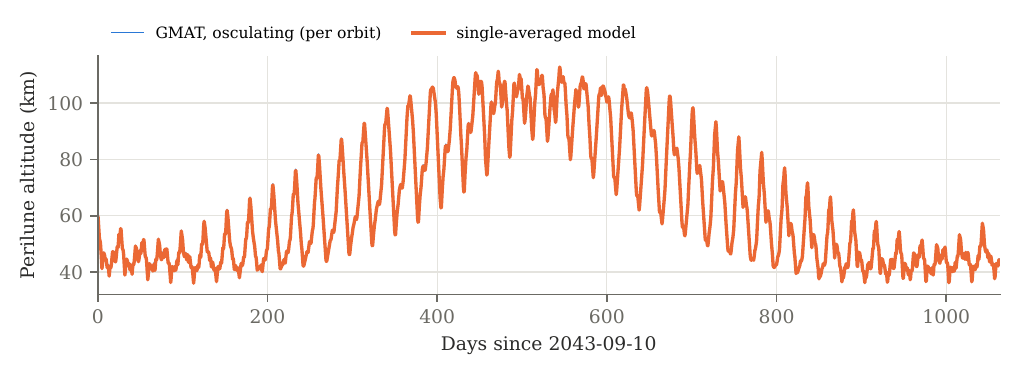}
\caption{Perilune altitude of case D5 ($20\times20$ field and Earth, $\Omega_0=90^\circ$): osculating values at each perilune passage from GMAT and the single-averaged model.}
\label{fig:savg_valid}
\end{figure}

Table~\ref{tab:savg_valid} compares the single-averaged model with GMAT propagations from the same initial state for several force models. The model reproduces the lowest altitude over three lunar years within 0.3~km, including the effect of the tesseral terms and of the Earth. An earlier version of the model, in which the field rotated about an axis $0.6^\circ$ from the true spin axis, differed by 1--5~km; the accuracy of the comparison depends on using exactly the rotation model of the propagator. Figure~\ref{fig:savg_valid} compares the perilune history of case D5 over the whole mission. With the $50\times50$ field the same model reproduces the lowest altitude of all twenty main GMAT runs within 0.5~km (Section~\ref{sec:sim_scan}).

\begin{table}[htbp]
\centering
\small
\caption{Lowest altitude over three lunar years: single-averaged model versus GMAT, same initial state ($\Omega_0=90^\circ$ case of Table~\ref{tab:sim_decomp}).}
\label{tab:savg_valid}
\begin{tabular}{llcc}
\toprule
Case & Force model & Averaged model (km) & GMAT (km) \\
\midrule
D1 & LP165P $20\times0$ & 53.8 & 53.9 \\
D2 & $20\times2$ & 38.5 & 38.8 \\
D3 & $20\times3$ & 34.2 & 34.4 \\
D4 & $20\times20$ & 29.8 & 29.8 \\
D5 & $20\times20$ + Earth & 36.1 & 35.9 \\
\bottomrule
\end{tabular}
\end{table}

\subsection{Convergence of the zonal series}
\label{sec:savg_zonal}
\begin{figure}[htbp]
\centering
\includegraphics[width=0.55\textwidth]{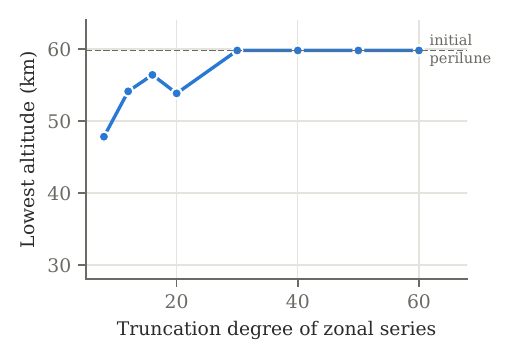}
\caption{Lowest altitude of the zonal-only problem over three lunar years versus the truncation degree of the zonal series (data of Table~\ref{tab:savg_conv}).}
\label{fig:zonal_conv}
\end{figure}

At $a=1868$~km the ratio $R_M/a=0.930$ decays slowly with degree, and the zonal coefficients of the Moon do not decrease fast enough to compensate. Table~\ref{tab:savg_conv} and Fig.~\ref{fig:zonal_conv} show the lowest altitude of the zonal-only problem as a function of the truncation degree. A model truncated at $J_8$ underestimates the lowest altitude by 12~km; truncation at degree 20 still differs by 6~km from the converged value, which is reached at degree 30. This is consistent with the recommendation of high-degree fields for low lunar orbits \cite{SanJuan2020}, and it quantifies the error of a low-degree design model for the present orbit. With the converged zonal series the eccentricity of the nominal orbit never exceeds its initial value: the start at $\omega_0=90^\circ$ lies at the maximum of $e$ on the $(e,\omega)$ contour, so that the initial perilune is also the lowest perilune of the zonal problem. With the Earth added, the lowest perilune of the mean motion is up to 2~km lower, depending on the inclination (58.6~km at the nominal $80.5^\circ$, Table~\ref{tab:thy_window}). The lowest altitude converges at degree~30, but the shape of the slow motion does not: the slow eccentricity vector rotates about a frozen center on the axis $\omega=90^\circ$. This point is the eccentricity vector of the frozen orbit, the equilibrium of the slow motion \cite{Cutting1978,ElipeLara2003}. We call it the frozen center because the orbits studied here are not frozen: their slow eccentricity vector rotates about this point, and whether that rotation encloses the origin decides whether the periapsis circulates (Section~\ref{sec:sim_circ}). The term describes this role and does not imply a nonlinear stability property. Its eccentricity $e_c$ (zonal terms only), at $i=80.5^\circ$ about the lunar equator, is 0.015 with $J_2$--$J_{20}$, 0.018 with $J_2$--$J_{30}$ and 0.023--0.024 from degree~50 to degree~100 (Fig.~\ref{fig:slow_centre}). Since the mean eccentricity vector of the nominal orbit passes the origin at a distance of about $|e_0-2e_c|$ in a circular approximation, its minimum and whether the mean periapsis circulates depend on degrees up to about~50 (Section~\ref{sec:sim_circ}).

\begin{table}[htbp]
\centering
\small
\caption{Lowest altitude of the zonal-only problem versus truncation degree (single-averaged model, three lunar years, initial state of case D1).}
\label{tab:savg_conv}
\begin{tabular}{lcccccccc}
\toprule
Truncation & $J_8$ & $J_{12}$ & $J_{16}$ & $J_{20}$ & $J_{30}$ & $J_{40}$ & $J_{50}$ & $J_{60}$ \\
\midrule
Lowest altitude (km) & 47.8 & 54.1 & 56.4 & 53.8 & 59.8 & 59.8 & 59.8 & 59.8 \\
\bottomrule
\end{tabular}
\end{table}

\section{Forced eccentricity from the nonzonal field}
\label{sec:forced}

\subsection{Derivation}
\label{sec:forced_deriv}
\paragraph{Selection rule} In Kaula's form \cite{Kaula1966} a term of degree $l$, order $m$ and indices $(p,q)$ of the disturbing function is
\begin{equation}
R_{lmpq}=\frac{\mu R_M^l}{a^{l+1}}\,F_{lmp}(i)\,G_{lpq}(e)\,J_{lm}\cos\!\big[(l-2p)\omega+(l-2p+q)M+m\psi-\chi_{lm}\big],
\label{eq:forced_kaula}
\end{equation}
with unnormalized coefficients, $J_{lm}=(C_{lm}^2+S_{lm}^2)^{1/2}$ and a phase $\chi_{lm}$ fixed by $C_{lm},S_{lm}$ and the parity of $l-m$. Averaging over $M$ keeps $q=-(l-2p)$. Since $G_{lpq}(e)=O(e^{|q|})$, the terms that are linear in $e$ have $l-2p=\pm1$, i.e.\ odd $l$ and $p=(l\mp1)/2$, with
\begin{equation}
\begin{gathered}
G_{l,p,\mp1}(e)=g_{lp}\,e\,\big[1+O(e^2)\big];\\
g_{3,1}=g_{3,2}=1,\qquad G_{3,1,-1}=G_{3,2,1}=e(1-e^2)^{-5/2}.
\end{gathered}
\end{equation}
Terms with $l-2p=0$ (even $l$, including $C_{22}$) do not depend on $\omega$ and leave $e$ unchanged, and terms with $|l-2p|\ge2$ are of order $e^2$ or smaller. For each odd $l$ and each $m\ge1$ the relevant part of the singly averaged potential is therefore
\begin{equation}
\begin{gathered}
\bar R_{lm}=e\,\big[K_{lm}^{+}\cos(\omega+m\psi-\chi_{lm})+K_{lm}^{-}\cos(-\omega+m\psi-\chi_{lm})\big],\\
K_{lm}^{\pm}=\frac{\mu R_M^l}{a^{l+1}}\,F_{lmp_\pm}(i)\,g_{lp_\pm}\,J_{lm},
\end{gathered}
\label{eq:forced_R}
\end{equation}
with $p_\pm=(l\mp1)/2$.

\paragraph{Complex eccentricity} Lagrange's equations give, to leading order in $e$,
\begin{equation}
\dot e=-\frac{\sqrt{1-e^2}}{na^2e}\frac{\partial\bar R}{\partial\omega},\qquad
e\dot\omega=\frac{\sqrt{1-e^2}}{na^2}\frac{\partial\bar R}{\partial e}-\frac{e\cot i}{na^2\sqrt{1-e^2}}\frac{\partial\bar R}{\partial i}.
\end{equation}
The last term is of order $e^2$ for Eq.~\eqref{eq:forced_R} and is dropped; this requires $\cot i$ to be bounded, i.e.\ an orbit away from the equatorial plane, as here ($\cot i=0.167$). With $\varepsilon=\xi+\mathrm i\eta=e\,\mathrm e^{\mathrm i\omega}$ and $\dot\varepsilon=(\dot e+\mathrm i e\dot\omega)\mathrm e^{\mathrm i\omega}$, the term with $K^{+}$ gives $\dot e+\mathrm i e\dot\omega=\mathrm i\kappa^{+}\mathrm e^{-\mathrm i(\omega+m\psi-\chi)}$ and the term with $K^{-}$ gives $\mathrm i\kappa^{-}\mathrm e^{\mathrm i(-\omega+m\psi-\chi)}$, where $\kappa^\pm=\sqrt{1-e^2}\,K^\pm/(na^2)$. Multiplying by $\mathrm e^{\mathrm i\omega}$,
\begin{equation}
\dot\varepsilon=\mathrm i\Big[\kappa_{lm}^{+}\,\mathrm e^{-\mathrm i(m\psi-\chi_{lm})}+\kappa_{lm}^{-}\,\mathrm e^{\mathrm i(m\psi-\chi_{lm})}\Big].
\label{eq:forced_epsdot}
\end{equation}
Two properties follow. To leading order in $e$, the forcing contains neither $\omega$ nor $e$: it is the sum of two counter-rotating components whose coefficients are independent of the eccentricity vector and whose phases advance with $m\psi$. It does not depend on where the perilune is, and it acts as strongly on a nearly circular orbit as on an eccentric one.

\paragraph{Forced solution} The slow zonal and Earth terms turn the eccentricity vector at the apsidal rate $\dot\omega_{\rm sec}$ ($\approx-0.39$~deg/day at the nominal elements; this is a local rate, whereas the drift of $\omega$ averaged over the GMAT runs that circulate is $-0.34$ to $-0.35$~deg/day, Table~\ref{tab:req_nominal}, and Section~\ref{sec:req} uses the rounded range $-0.35$ to $-0.4$~deg/day for planning). Linearizing about the slow motion, $\dot\varepsilon=\mathrm i\dot\omega_{\rm sec}\varepsilon+\dot\varepsilon_{lm}$, with $\dot\varepsilon_{lm}$ from Eq.~\eqref{eq:forced_epsdot} and $\psi=\psi_0+\nu_\psi t$, $\nu_\psi\simeq\dot\Omega-\dot\theta$, the particular solution is
\begin{equation}
\begin{split}
\varepsilon&=\bar\varepsilon+\delta\varepsilon(\psi),\\
\delta\varepsilon(\psi)&=\sum_{l\ \mathrm{odd}}\sum_{m\ge1}\Bigl[-\frac{\kappa_{lm}^{+}\,\mathrm e^{-\mathrm i(m\psi-\chi_{lm})}}{m\nu_\psi+\dot\omega_{\rm sec}}+\frac{\kappa_{lm}^{-}\,\mathrm e^{\mathrm i(m\psi-\chi_{lm})}}{m\nu_\psi-\dot\omega_{\rm sec}}\Bigr],
\end{split}
\label{eq:forced_sol}
\end{equation}
where $\bar\varepsilon$ evolves on the slow contour of the zonal problem. The scalar apsidal-rate term $\mathrm i\dot\omega_{\rm sec}\varepsilon$ is a local approximation to the response about the slow trajectory, not a global model of the slow motion, which rotates about a non-zero center; the general local equation involves the Jacobian of the slow vector field. This is one reason why the amplitude formula works well while the two-step decomposition can still overestimate the late lowest perilune (Section~\ref{sec:forced_node}), an error that remains with the nonsingular elements. Each pair $(l,m)$ traces an ellipse with period $2\pi/(m|\nu_\psi|)\approx27.0/m$~days and semi-axes $|\alpha^+_{lm}|+|\alpha^-_{lm}|$ and $\big||\alpha^+_{lm}|-|\alpha^-_{lm}|\big|$, where $\alpha^\pm_{lm}=\kappa^\pm_{lm}/(m\nu_\psi\pm\dot\omega_{\rm sec})$. Since $|\dot\omega_{\rm sec}|\ll|\nu_\psi|$ (0.39 against 13.3~deg/day), the apsidal rate changes the two circular components by about $\mp3\%/m$ in opposite directions; the net change of the largest excursion is $-1.4\%$, $+0.6\%$ and $-0.2\%$ for $m=1,2,3$. Using $\mu/a^3=n^2$ and neglecting $\dot\omega_{\rm sec}$, the largest excursion of one pair is, to leading order in $e$,
\begin{equation}
|\delta\varepsilon_{lm}|_{\max}=\frac{n}{m|\nu_\psi|}\Big(\frac{R_M}{a}\Big)^{l}\big(|F_{lmp_+}g_{lp_+}|+|F_{lmp_-}g_{lp_-}|\big)\,J_{lm}\,\big[1+O(e^2)\big].
\label{eq:forced_amp}
\end{equation}
For $l=3$ the eccentricity functions are known exactly, $G_{3,p,\mp1}=e(1-e^2)^{-5/2}$, so that $g_{3,1}=g_{3,2}=1$; the $O(e^2)$ correction differs between the $\dot e$ and $e\dot\omega$ components, $(1-e^2)^{-2}$ and $(1+4e^2)(1-e^2)^{-3}$, and amounts to less than 1\% here. Hence
\begin{equation}
|\delta\varepsilon_{3m}|_{\max}\simeq\frac{n}{m|\nu_\psi|}\Big(\frac{R_M}{a}\Big)^{3}\big(|F_{3m1}|+|F_{3m2}|\big)J_{3m},\qquad |\nu_\psi|\approx\dot\theta .
\label{eq:forced_amp3}
\end{equation}
The factor $n/|\nu_\psi|\approx322$ is essentially the ratio of the orbital to the lunar rotation rate. The slow rotation of the Moon thus amplifies weak nonzonal terms into a large eccentricity response. Equations~\eqref{eq:forced_epsdot}--\eqref{eq:forced_sol} are the monthly eccentricity-vector correction of the odd-degree tesseral and sectoral harmonics that underlies the translation theorem \cite{Yarndley2025,Yarndley2026}, and the amplification is the one analyzed by Lara \cite{Lara2025}. What the Kaula form adds here is the explicit amplitude of each $(l,m)$ pair, Eq.~\eqref{eq:forced_amp}, which allows the perilune margin of a given orbit to be evaluated without propagation.

\paragraph{Degree three} For $l=3$ the inclination functions are
\begin{align}
F_{311}&=\tfrac{15}{16}s^2(1+3c)-\tfrac34(1+c), & F_{312}&=\tfrac{15}{16}s^2(1-3c)-\tfrac34(1-c),\nonumber\\
F_{321}&=\tfrac{15}{8}s(1-2c-3c^2), & F_{322}&=-\tfrac{15}{8}s(1+2c-3c^2),\label{eq:forced_F3}\\
F_{331}&=\tfrac{45}{8}s^2(1+c), & F_{332}&=\tfrac{45}{8}s^2(1-c),\nonumber
\end{align}
with $s=\sin i$ and $c=\cos i$. Table~\ref{tab:forced_closed} evaluates Eq.~\eqref{eq:forced_amp3} for the nominal orbit. The closed form agrees with the numerical Fourier analysis of the singly averaged rates at fixed elements within 2\% for all three orders. In single-averaged propagations of each pair on the converged zonal problem ($\dot\omega_{\rm sec}=-0.19$~deg/day without the Earth), the forced amplitude agrees with the closed form within 1.3\% for $m=2,3$ and is 3.8\% larger for $m=1$ (4.5\% larger than the closed form with the apsidal correction, which moves the $m=1$ value the other way). The difference lies in the retrograde ($\kappa^-$) component, which is 8--15\% above the closed form and drifts with $e$ and $\omega$; because $F_{312}=-0.166$ results from the cancellation of two terms of 0.46 and 0.63, this component is sensitive to small corrections outside the first-order theory, which we have not identified. Relative to propagation, the closed form is therefore accurate to about 5\%. The $m=1$ pair dominates because $J_{31}$ is the largest low-degree tesseral coefficient of the Moon and because the $1/m$ factor favors low orders.

\begin{table}[htbp]
\centering
\small
\caption{Forced eccentricity of the degree-3 terms for the nominal orbit ($a=1868$~km, $e=0.0375$, $i=80.5^\circ$), as $a|\delta\varepsilon|_{\max}$ in km: closed form~\eqref{eq:forced_amp3}, with the apsidal-rate correction of Eq.~\eqref{eq:forced_sol} for $\dot\omega_{\rm sec}=-0.19$~deg/day, numerical Fourier analysis of the singly averaged rates at fixed elements, and single-averaged propagation of each pair with the zonal series to degree 30 (400 days).}
\label{tab:forced_closed}
\setlength{\tabcolsep}{4pt}
\begin{tabular}{lccccccc}
\toprule
$m$ & $F_{3m p_+}$ & $F_{3m p_-}$ & $J_{3m}$ & Eq.~\eqref{eq:forced_amp3} & $+\dot\omega_{\rm sec}$ & Fourier & Prop. \\
\midrule
1 & $0.490$ & $-0.166$ & $2.90\times10^{-5}$ & 9.22 & 9.16 & 9.34 & 9.57 \\
2 & $1.088$ & $-2.309$ & $5.13\times10^{-6}$ & 4.22 & 4.23 & 4.25 & 4.27 \\
3 & $6.375$ & $4.569$ & $1.73\times10^{-6}$ & 3.06 & 3.06 & 3.07 & 3.06 \\
\bottomrule
\end{tabular}
\end{table}

\paragraph{Scaling} Equation~\eqref{eq:forced_amp} gives the design trends directly. To leading order in $e$, at fixed semimajor axis and inclination, the forced amplitude is independent of the eccentricity vector; it grows as $n(R_M/a)^l$ when the orbit is lowered; it depends on the inclination only through $F_{lmp}(i)$; and its relative effect on the perilune, $a|\delta\varepsilon|/(a e)$, is largest for nearly circular orbits. For low lunar orbits with $e\lesssim0.05$ the tesseral forced eccentricity is therefore of the same order as the eccentricity itself.

\subsection{Consequences for the perilune}
\label{sec:forced_perilune}
\begin{figure}[htbp]
\centering
\includegraphics[width=0.98\textwidth]{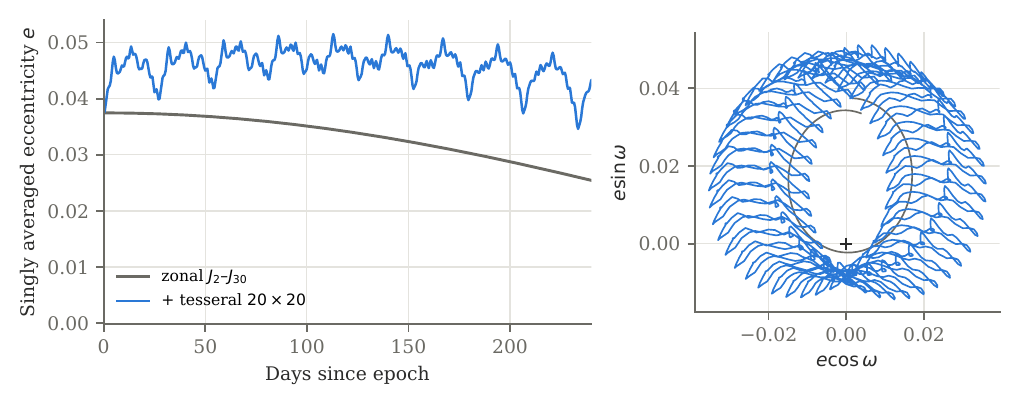}
\caption{Forced eccentricity. Left: singly averaged eccentricity (which keeps the monthly terms) over the first 240 days for the converged zonal problem and with the $20\times20$ tesseral field added; the monthly oscillation is the forced term of Eq.~\eqref{eq:forced_sol}. Right: eccentricity vector over three lunar years for the same two cases.}
\label{fig:forced_ecc}
\end{figure}

With $\delta r_p=-a\,\delta e$ and the first-order decomposition $\varepsilon=\bar\varepsilon+\delta\varepsilon$, the triangle inequality gives the lower bound
\begin{equation}
h_{p,\min}\ge a\big(1-|\bar\varepsilon|_{\max}-|\delta\varepsilon|_{\max}\big)-R_{\rm alt} ,
\label{eq:forced_bound}
\end{equation}
where $|\bar\varepsilon|_{\max}$ is the largest eccentricity on the zonal contour. The bound is reached only if the forced vector aligns with the slow one at the time when $|\bar\varepsilon|$ is largest; since the forced vector rotates monthly while $|\bar\varepsilon|$ changes over the circulation, full alignment is not guaranteed and the bound is conservative. Table~\ref{tab:forced_amp} gives $|\delta\varepsilon|_{\max}$ for the nominal orbit, evaluated by Fourier analysis in $\psi$ of the singly averaged rates~\eqref{eq:savg_avg} (this avoids truncating the inclination and eccentricity functions). The degree-3 terms alone give 13.2~km, the $m=1$ terms up to degree 20 give 18.9~km, and the full $20\times20$ field gives 25.3~km. Most of the forced amplitude is captured by degree~10 (23.8~km), with smaller changes beyond degree~20 (less than 0.3~km). $C_{22}$ contributes less than 1~km. Figure~\ref{fig:forced_ecc} shows the forced oscillation in a single-averaged propagation.

\begin{table}[htbp]
\centering
\small
\caption{Forced eccentricity amplitude for the nominal orbit ($a=1868$~km, $e=0.0375$, $i=80.5^\circ$) and its perilune equivalent $a|\delta\varepsilon|_{\max}$, compared with the decrease of the lowest altitude in single-averaged propagations on the converged zonal problem (59.8~km). The propagations start from the same osculating state ($\Omega_0=90^\circ$); because the mean eccentricity then also shifts by $\delta\varepsilon(\psi_0)$, Eq.~\eqref{eq:forced_mean0}, the decrease can exceed $a|\delta\varepsilon|_{\max}$ (here by up to 10\%).}
\label{tab:forced_amp}
\begin{tabular}{lccc}
\toprule
Terms & $|\delta\varepsilon|_{\max}$ & $a|\delta\varepsilon|_{\max}$ (km) & Propagated decrease (km) \\
\midrule
$C_{22}$ & 0.0005 & 0.9 & 0.02 \\
$C_{31},S_{31}$ & 0.0050 & 9.3 & 7.1 \\
$C_{32},S_{32}$ & 0.0023 & 4.3 & --- \\
$C_{33},S_{33}$ & 0.0016 & 3.1 & --- \\
all $l=3$ & 0.0071 & 13.2 & 14.5 \\
all $m=1$, $l\le20$ & 0.0101 & 18.9 & 11.9 \\
full $20\times20$ & 0.0136 & 25.3 & 26.2 \\
\midrule
full, degree $\le10$ & 0.0127 & 23.8 & --- \\
full, degree $\le30$ & 0.0135 & 25.2 & --- \\
full, degree $\le50$ & 0.0134 & 25.1 & --- \\
\bottomrule
\end{tabular}
\end{table}

\subsection{Mean and osculating initial eccentricity: origin of the node dependence}
\label{sec:forced_node}
The injection fixes the osculating eccentricity vector $\varepsilon_0$, not the mean one; the importance of this distinction for initializing low lunar orbits was emphasized by Nie and Gurfil \cite{NieGurfil2018} and Lara \cite{Lara2025}. By Eq.~\eqref{eq:forced_sol} the mean vector that governs the slow motion is
\begin{equation}
\bar\varepsilon_0=\varepsilon_0-\delta\varepsilon(\psi_0),
\label{eq:forced_mean0}
\end{equation}
which depends on the initial phase $\psi_0=\Omega_0-\theta_0$. The lowest altitude of a run started from given osculating elements is therefore
\begin{equation}
h_{p,\min}(\psi_0)\simeq a\Big(1-\max_{\psi}\big|\varepsilon_0-\delta\varepsilon(\psi_0)+\delta\varepsilon(\psi)\big|\Big)-R_{\rm alt} ,
\label{eq:forced_hmin}
\end{equation}
which varies with the initial node through $\delta\varepsilon(\psi_0)$. The $m=1$ terms dominate $\delta\varepsilon$ and have period $360^\circ$ in $\psi_0$, which explains a $360^\circ$ component of the node dependence that the first-order inclination law~\eqref{eq:thy_dI} cannot produce.

\begin{figure}[htbp]
\centering
\includegraphics[width=0.98\textwidth]{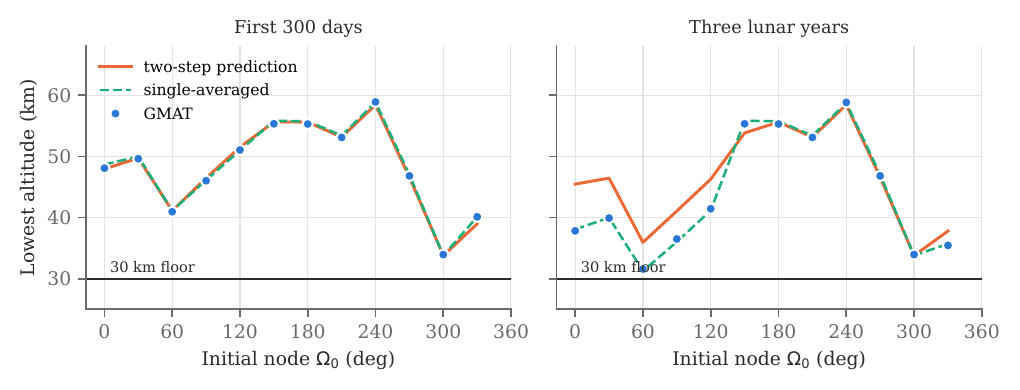}
\caption{Lowest altitude versus initial node over the first 300 days (left) and over three lunar years (right): GMAT propagations (T000--T330) and the prediction of Eq.~\eqref{eq:forced_twostep} and the full single-averaged model, all with the $50\times50$ field and no fitted parameter.}
\label{fig:raan_scan}
\end{figure}

Equation~\eqref{eq:forced_hmin} assumes that the slow vector stays near its initial value. Over a full circulation the slow motion itself can bring $|\bar\varepsilon|$ back to a maximum, and the alignment with the forced term then occurs late in the mission. We therefore evaluate the lowest perilune in two steps: (i) propagate the slow vector $\bar\varepsilon(t)$ from $\bar\varepsilon_0$ with the zonal series and the Earth only, which costs seconds; (ii) superpose the forced term at the actual phase,
\begin{equation}
h_{p,\min}=\min_t\Big[a\big(1-\big|\bar\varepsilon(t)+\delta\varepsilon\big(\psi(t)\big)\big|\big)\Big]-R_{\rm alt} ,\qquad \psi(t)=\psi_0+\int_0^t\big(\dot\Omega-\dot\theta\big)\,\dd t .
\label{eq:forced_twostep}
\end{equation}
Altitudes in all comparisons with GMAT are measured from the GMAT lunar radius of 1738.2~km, also in Eqs.~\eqref{eq:forced_hmin} and~\eqref{eq:forced_twostep}, whereas the field is referred to the LP165P radius of 1738.0~km. The initial vector $\varepsilon_0$ in Eq.~\eqref{eq:forced_mean0} is taken directly from the osculating GMAT elements at the epoch; the orbital-period terms are not removed first, and their effect is contained in the agreement reported below. Step (i) uses the nonsingular elements of Section~\ref{sec:savg_model}, the zonal terms to degree~50 and the Earth, and step (ii) the forced term of all tesseral terms to degree and order~50, matching the $50\times50$ field of the GMAT runs. No parameter of the model is fitted to the GMAT results. Two modeling choices were, however, corrected after earlier comparisons with GMAT: the rotation axis of the field (Section~\ref{sec:sim_frame}) and the truncation of the field, which must match the propagator (below); both are properties of the propagator that the model must reproduce, not adjustable parameters. With these, Eq.~\eqref{eq:forced_twostep} reproduces the lowest altitude of the twelve-node GMAT scan of Section~\ref{sec:sim_scan} over the first 300 days with a largest difference of 1.1~km (rms 0.4~km) (Table~\ref{tab:forced_scan}, Fig.~\ref{fig:raan_scan}). In this phase the node dependence, a spread of 25~km, is predominantly explained by the forced correction evaluated at the insertion phase. Later in the mission the largest errors of Eq.~\eqref{eq:forced_twostep} (prediction minus GMAT) are overestimates of the lowest altitude, 7.7~km at the end of the mission and up to 9.3~km at intermediate horizons, in the runs in which the mean eccentricity vector passes close to the origin (Section~\ref{sec:sim_circ}), where the separation of slow and forced motion assumed by Eq.~\eqref{eq:forced_twostep} is least accurate. Averaging the slow rates over the forced cycle, i.e.\ including the second-order coupling of the two motions, reduces this difference at T000 from 7.7 to 5.5~km; the remainder has not been identified. The single-averaged model integrated in full, without the two-step decomposition, does not have this limitation: over the twenty GMAT runs of Sections~\ref{sec:sim_scan} and~\ref{sec:sim_target} it reproduces the three-year lowest altitude with an rms difference of 0.27~km and a largest difference of 0.48~km, and the lowest altitudes of days 0--300, 300--700 and 700--1063 equally well.

These comparisons require the same truncation of the field in the model and in the propagator. An earlier version of the analysis used $J_2$--$J_{30}$ in the model against $J_2$--$J_{20}$ in a $20\times20$ GMAT field and showed an apparent late drift of up to 9~km, which disappears with matched truncation; with the $(e,\omega)$ elements the model also differed by up to 3.8~km from GMAT in runs whose slow eccentricity passes close to zero, a difference that the nonsingular elements remove (0.2~km for the same run).

\begin{figure}[htbp]
\centering
\includegraphics[width=0.55\textwidth]{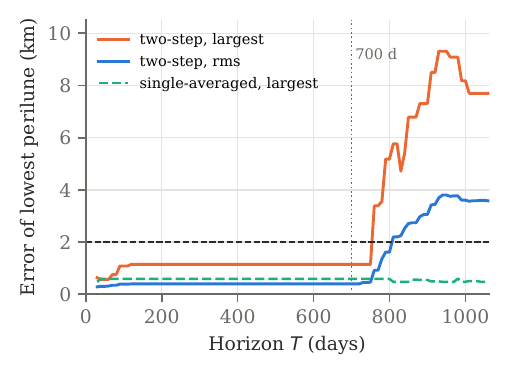}
\caption{Absolute error in the lowest perilune reached up to day $T$ over the twenty GMAT runs ($50\times50$ field) T000--T330, M000--M300, H060 and H300: largest and rms absolute error of Eq.~\eqref{eq:forced_twostep}, and largest absolute error of the full single-averaged model. The largest two-step error over all horizons is 9.3~km ($T\approx930$~days) and 7.7~km at the mission end; the full single-averaged model stays within 0.6~km at all horizons and within 0.5~km at the mission end. Dashed: 2~km; dotted: $T=700$~days.}
\label{fig:validity}
\end{figure}

For design, the relevant quantity is the lowest perilune reached up to a time $T$. Over the twenty GMAT runs, the error of Eq.~\eqref{eq:forced_twostep} in this quantity (Fig.~\ref{fig:validity}) stays within 1.2~km (rms 0.4~km) up to $T\approx700$~days, reaches its largest value over all horizons, 9.3~km, at $T\approx930$~days, and is 7.7~km (rms 3.6~km) at the end of the mission; the largest errors are overestimates, while at some nodes the prediction is low by up to 1.5~km (T150). Over the same horizons the full single-averaged model stays within 0.6~km at every horizon. Within the tested set, Eq.~\eqref{eq:forced_twostep} is therefore a design estimate accurate to about 1~km for the first 23 lunar months; for the full mission the single-averaged model, at a few minutes per three-year run with the $50\times50$ field, is the tool for design and verification.

\begin{table}[htbp]
\centering
\small
\caption{Lowest altitude (km) versus initial node over the first 300 days and over three lunar years: two-step prediction of Eq.~\eqref{eq:forced_twostep}, full single-averaged model and GMAT (runs T000--T330), all with the $50\times50$ field.}
\label{tab:forced_scan}
\setlength{\tabcolsep}{3pt}
\begin{tabular}{lcccccccccccc}
\toprule
$\Omega_0$ (deg) & 0 & 30 & 60 & 90 & 120 & 150 & 180 & 210 & 240 & 270 & 300 & 330 \\
\midrule
\multicolumn{13}{l}{\emph{Days 0--300}}\\
Two-step & 47.9 & 49.7 & 41.1 & 46.5 & 51.5 & 55.6 & 55.6 & 53.1 & 58.4 & 46.6 & 33.8 & 39.0 \\
Model & 48.7 & 50.0 & 41.1 & 46.1 & 51.0 & 55.8 & 55.7 & 53.4 & 58.8 & 47.0 & 33.8 & 40.1 \\
GMAT & 48.1 & 49.6 & 40.9 & 46.0 & 51.0 & 55.3 & 55.3 & 53.1 & 58.9 & 46.8 & 33.9 & 40.1 \\
\multicolumn{13}{l}{\emph{Three lunar years}}\\
Two-step & 45.5 & 46.4 & 35.9 & 41.1 & 46.3 & 53.8 & 55.6 & 53.1 & 58.4 & 46.6 & 33.8 & 37.8 \\
Model & 38.1 & 40.0 & 31.3 & 36.0 & 41.1 & 55.8 & 55.7 & 53.4 & 58.7 & 47.0 & 33.8 & 35.7 \\
GMAT & 37.8 & 39.9 & 31.6 & 36.5 & 41.4 & 55.3 & 55.3 & 53.1 & 58.8 & 46.8 & 33.9 & 35.4 \\
\bottomrule
\end{tabular}
\end{table}

\subsection{Design rule: targeting the mean eccentricity}
\label{sec:forced_rule}
Equation~\eqref{eq:forced_mean0} also gives the remedy. If the injection targets the osculating vector
\begin{equation}
\varepsilon_0=\bar\varepsilon^{\ast}+\delta\varepsilon(\psi_0),
\label{eq:forced_target}
\end{equation}
where $\bar\varepsilon^{\ast}$ is the design mean eccentricity vector, the principal node dependence of the initial mean eccentricity vector is removed at first order (the slow motion that follows still depends on the node through the zonal and Earth terms), and Eq.~\eqref{eq:forced_bound} with $|\bar\varepsilon|_{\max}=|\bar\varepsilon^{\ast}|$ gives a conditional lower bound within the first-order model: it holds as long as the slow motion does not raise $|\bar\varepsilon|$ above its design value. On the converged zonal contour the start at $\omega_0=90^\circ$ is the maximum of $e$, but with the Earth included the largest slow eccentricity of a targeted orbit has not been bounded in general. For the nominal orbit, $a|\delta\varepsilon|_{\max}=25.1$~km with the $50\times50$ field (25.3~km with $20\times20$, Table~\ref{tab:forced_amp}), so the conditional bound is 34.7~km for a 60~km mean perilune (59.8~km above the GMAT radius of 1738.2~km) with the zonal terms alone; these numbers are illustrative estimates under the stated assumptions, not a safety floor. The bound is conservative because the forced and slow vectors do not reach full alignment at the moment of largest $|\bar\varepsilon|$; for the targeted nominal orbit at six nodes GMAT ($50\times50$) gives 50.2--50.9~km over the first 300 days, as predicted by Eq.~\eqref{eq:forced_twostep}, and 39.9--50.9~km over three lunar years (Section~\ref{sec:sim_target}). The required correction is small: $|\delta\varepsilon(\psi_0)|\le0.014$, i.e.\ a change of at most 25~km in the apsidal altitudes and a rotation of the line of apsides by up to about $20^\circ$, which can be built into the insertion sequence.

\section{Full-force verification}
\label{sec:sim}

\subsection{Setup}
\label{sec:sim_setup}
All propagations start at 2043-09-10 00:00:00~UTC. The baseline runs (T, D, S090, E, F, G) start from the nominal osculating elements of Table~\ref{tab:req_nominal}; the targeted runs (M, H) start from osculating elements adjusted by Eq.~\eqref{eq:forced_target} (Table~\ref{tab:sim_target}), and runs I5--I8 use the corrected inclination of Eq.~\eqref{eq:thy_dI}. The initial inclination $i_0=80.5^\circ$ is measured from the axis fitted to case D1 for the T, E, F, G, M and H runs and from the IAU equator of the epoch for the intermediate set (D1--D6, I5--I8); Section~\ref{sec:sim_frame} gives the resulting inclinations about the effective precession axis. The force model of Table~\ref{tab:bg_forces} is the baseline. The main runs (T, M, H) use the LP165P field to degree and order~50 and were also run with a $20\times20$ field for comparison (Section~\ref{sec:sim_circ}); the decomposition and term-group runs, which isolate the mechanism, use the $20\times20$ field with the force terms reduced as listed in Table~\ref{tab:sim_runs}. All runs last $36\times29.53=1063$~days. Each run records the osculating state at every periapsis and apoapsis passage, which gives the extreme altitudes of each revolution directly. Altitudes are reported above the GMAT lunar radius of 1738.2~km, so the nominal orbit starts at $59.8\times199.8$~km. The runs are listed in Table~\ref{tab:sim_runs}: a scan of the initial node at the nominal inclination, targeted runs, a force-model decomposition at one node, runs that isolate groups of terms at four nodes, and runs with the inclination corrected by Eq.~\eqref{eq:thy_dI}. A three-year run takes about eight minutes on a desktop computer with the $20\times20$ field and correspondingly longer with the $50\times50$ field.

\begin{table}[htbp]
\centering
\small
\caption{GMAT runs.}
\label{tab:sim_runs}
\begin{tabular}{@{}p{0.19\textwidth}p{0.19\textwidth}p{0.52\textwidth}@{}}
\toprule
Group & Runs & Purpose \\
\midrule
Node scan & T000--T330 & $\Omega_0=0^\circ,30^\circ,\dots,330^\circ$, full force model, $50\times50$ field (repeated with $20\times20$), $i_0=80.5^\circ$ about the fitted axis (Section~\ref{sec:sim_frame}) \\
Targeting & M000--M300 & $\Omega_0=0^\circ,60^\circ,\dots,300^\circ$, mean $60\times200$~km orbit with targeted mean eccentricity, $50\times50$ (repeated with $20\times20$) (Table~\ref{tab:sim_target}) \\
 & H060, H300 & as M060, M300 with a 70~km mean perilune \\
Decomposition & D1--D6, S090 & $\Omega_0=90^\circ$; force terms added one at a time, up to $20\times20$ (Table~\ref{tab:sim_decomp}) \\
Term groups & E1, E2 & $J_2$ only; $J_2$ and $C_{22}$ only \\
 & F000--F270 & zonal terms and all $m\le2$ terms to degree 20, no third body \\
 & G000--G270 & zonal terms to degree 20 and the Earth, no tesseral terms \\
Inclination law & I5--I8 & $\Omega_0=0^\circ,90^\circ,180^\circ,270^\circ$, $20\times20$, with $i_0=80.5^\circ+0.606^\circ\cos2\psi_0$ \\
\bottomrule
\end{tabular}
\end{table}

\subsection{Definition of the initial inclination in GMAT}
\label{sec:sim_frame}
The Moon-centered inertial frame of GMAT (\texttt{BodyInertial} axes) is fixed to the lunar equator at J2000. Because the lunar pole precesses, this equator is not the equator that the gravity field rotates about in 2043. A first set of runs, whose initial elements were given in this frame, showed an apparent decrease of the inclination from $80.5^\circ$ to $75^\circ$ over three years, a perilune as low as 33~km, and a strong dependence on the initial node. These differences arose from specifying the initial elements relative to an unintended reference plane: the inclination with respect to the equator of 2043 was between $77.7^\circ$ and $83.3^\circ$, depending on $\Omega_0$.

To remove this effect we first identified, from a zonal-only run (case D1), the axis about which $\sqrt{1-e^2}\cos i$ is conserved, by fitting a slowly moving axis to the run. The fit leaves a residual of $1.5\times10^{-5}$; the fitted axis is $3.29^\circ$ from the GMAT $Z$ axis at the epoch, whereas the spin axis of the IAU model \cite{Archinal2011} is $2.78^\circ$ from it, and the two are $0.6^\circ$ apart. The runs T, E, F, G, M and H were built with $i_0=80.5^\circ$ about this fitted axis.

The difference was explained afterward. In our GMAT version and settings (GMAT R2026a, default DE405 file \texttt{leDE1941.405}), GMAT rotates the lunar field with the DE405 libration angles \cite{Standish1998}; a direct output of its body-fixed axes at eight epochs coincides with the DE405 principal-axis frame to better than $10^{-4}$ degree and lies within $0.03^\circ$ of the IAU spin axis. The fitted axis is therefore not the spin axis but an approximation, over half a nodal cycle, of the effective precession axis $\mathbf w$ of Eq.~\eqref{eq:bg_waxis}. With $\mathbf w$ built from the spin axis and the ecliptic pole, a fit of the coefficient of $\mathbf k$ to case D1 gives $-0.331$, consistent with $-\gamma/\dot\Omega=-0.331$, and conserves $\sqrt{1-e^2}\cos i$ to $3\times10^{-5}$. Because the fit covers only part of a nodal cycle, it is not unique: the fitted axis and $\mathbf w$ are $0.5^\circ$--$0.7^\circ$ apart, and the initial inclinations of the T runs range from $79.8^\circ$ to $81.2^\circ$ about $\mathbf w$ and from $79.9^\circ$ to $81.1^\circ$ about the lunar equator of the epoch, instead of being exactly $80.5^\circ$. Over this range the lowest perilune of the zonal and Earth problem changes by less than 2~km (Table~\ref{tab:thy_window}), and runs I5--I8 ($20\times20$) show that a change of the initial inclination by $0.6^\circ$ moves the lowest altitude by at most 0.2~km. The insertion-phase correction explains the dominant variation of the lowest perilune (Section~\ref{sec:sim_scan}); the contribution of the accompanying spread of the initial inclination to the near-critical slow motion has not been isolated with the $50\times50$ field. All predictions of Sections~\ref{sec:forced} and~\ref{sec:sim} start from the actual GMAT initial states and use the DE405 body axes, so they are not affected by the choice of reference axis. An intermediate set, defined with respect to the IAU equator of the epoch, contains the decomposition D1--D6 and the runs I5--I8; it is used only for comparisons at a common initial state. Inclinations in this section are measured from $\mathbf w$.

The node labels $\Omega_0$ of the two sets are referred to different planes and origins and are therefore not comparable. The physical variable is the insertion phase $\psi_0=\Omega_0-\theta_0$, the node of the initial orbit plane on the lunar equator measured from the prime meridian of the DE405 body frame. For the T, M and H runs $\psi_0=\Omega_0+85.0^\circ$ (within $0.1^\circ$), and for the D and I runs $\psi_0=\Omega_0-60.0^\circ$. Comparisons at nearby insertion phases show broadly consistent trends, although the cases are not matched in all initial elements and the phases differ by $5^\circ$ (all with the $20\times20$ field): D5 ($\psi_0=30^\circ$) reaches 35.9~km and T300 ($25^\circ$) 36.2~km; I5--I8 ($\psi_0=300^\circ,30^\circ,120^\circ,210^\circ$) reach 53.3, 35.8, 41.6 and 53.2~km, and T210, T300, T030 and T120 ($295^\circ,25^\circ,115^\circ,205^\circ$) 53.3, 36.2, 42.0 and 49.6~km; the last pair differs by 3.6~km. A strict comparison would convert the same Cartesian state with both element definitions, and the effect of the initial inclination on the near-critical circulation would need a controlled test with the $50\times50$ field. The node scan is labeled by $\Omega_0$ for continuity, and $\psi_0$ is given in Table~\ref{tab:sim_scan}.

The lesson is general. A propagator's inertial lunar frame may be tied to a reference epoch decades before the mission, and the spin axis of the epoch is not the axis about which a low orbit precesses. The initial state should be specified with the rotation model of the propagator, and the effective precession axis of Eq.~\eqref{eq:bg_waxis} is a convenient reference for the orbit plane, and any model used for prediction must rotate the field exactly as the propagator does: an axis error of $0.6^\circ$ changed our predicted lowest perilunes by up to 5~km.

\subsection{Force-model decomposition}
\label{sec:sim_decomp}
Table~\ref{tab:sim_decomp} lists the decomposition at $\Omega_0=90^\circ$ of the intermediate set ($\psi_0=30^\circ$). The zonal-only run D1 agrees with the single-averaged model and stays close to the initial perilune; the difference to the converged value (Table~\ref{tab:savg_conv}) is the truncation at degree~20. Adding the order-1 and order-2 terms (D2) lowers the perilune by 15~km, the order-3 terms (D3) by 4~km more, and the rest of the $20\times20$ field (D4) by another 5~km, to 29.8~km, below the floor. The Earth raises the lowest altitude by 6~km (D5); the Sun and radiation pressure change it by less than 0.1~km. The order-1 and order-2 group of D2 contains $C_{22}$, but the single-averaged experiments of Table~\ref{tab:forced_amp} show that nearly all of its effect comes from the $m=1,2$ terms of degree $l\ge3$. The term-group runs confirm this division. With $J_2$ and $C_{22}$ only (E2) the eccentricity stays between 0.0374 and 0.0381, exactly as with $J_2$ alone (E1), so $C_{22}$ does not move the eccentricity. With the zonal terms and the Earth but no tesseral terms (G000--G270) the lowest altitude is 54.1--59.8~km, within 6~km of the initial perilune at every node. With the zonal and all $m\le2$ terms but no Earth (F000--F270) it is 37.0--59.8~km, and Eq.~\eqref{eq:forced_twostep} restricted to the $m\le2$ terms predicts 39.1, 45.8, 59.3 and 48.3~km against the GMAT values 37.0, 42.6, 59.8 and 48.7~km, i.e.\ within 3.2~km.

\begin{table}[htbp]
\centering
\small
\caption{Force-model decomposition at $\Omega_0=90^\circ$ (GMAT, three lunar years).}
\label{tab:sim_decomp}
\begin{tabular}{llcccc}
\toprule
Case & Force model & Lowest (km) & Highest (km) & $e$ range & $\Delta\omega$ (deg) \\
\midrule
D1 & LP165P $20\times0$ & 53.9 & 207.0 & 0.006--0.041 & $-380$ \\
D2 & $20\times2$ & 38.8 & 222.6 & 0.001--0.049 & $-360$ \\
D3 & $20\times3$ & 34.4 & 226.7 & 0.002--0.051 & $-360$ \\
D4 & $20\times20$ & 29.8 & 231.0 & 0.002--0.054 & $-358$ \\
D5 & D4 + Earth & 35.9 & 225.0 & 0.009--0.051 & $-376$ \\
D6 & D5 + Sun & 35.9 & 225.0 & 0.009--0.051 & $-376$ \\
S090 & D6 + SRP & 36.0 & 225.0 & 0.009--0.050 & $-376$ \\
\bottomrule
\end{tabular}
\end{table}

\subsection{Node scan}
\label{sec:sim_scan}
Table~\ref{tab:sim_scan} summarizes the twelve-node scan with the $50\times50$ field. No run reaches the surface, and the perilune never falls below 30~km. The lowest altitude ranges from 31.6~km ($\Omega_0=60^\circ$) to 58.8~km ($240^\circ$), and the highest from 202~km to 229~km; the two deviations are nearly symmetric, as expected for a variation of $e$ at fixed $a$. A harmonic fit of the lowest altitude against $\Omega_0$ has a $360^\circ$ component of 11.3~km and a $180^\circ$ component of 4.0~km. Over the first 300 days Eq.~\eqref{eq:forced_twostep} reproduces the curve within 1.1~km at every node, and the full single-averaged model reproduces the mission-minimum perilune altitude within 0.5~km at every node (Table~\ref{tab:forced_scan}, Fig.~\ref{fig:raan_scan}); the node dependence is therefore a property of the dynamics and not of the frame (Section~\ref{sec:sim_frame}). The comparison supports the insertion-phase mechanism, although the contribution of the accompanying inclination spread has not been isolated with the $50\times50$ field. With the $20\times20$ field the lowest altitudes are higher by up to 8.2~km (T120), most of all at the nodes where the late perilune is lowest, and differ by less than 1~km where the lowest altitude is reached in the first months. The model with degree and order~70 and~100 gives the lowest altitude of T090 as 38.0 and 37.9~km, against 36.0~km at degree~50; for T060 and M120 the change from degree~50 to~70 is $+0.8$ and $+1.6$~km. The $50\times50$ results are therefore slightly conservative for the late perilune in the cases tested at higher degree, by up to about 2~km; this has not been checked at the other nodes.

\begin{table}[htbp]
\centering
\small
\caption{Node scan (GMAT, $50\times50$ field, three lunar years; values with the $20\times20$ field in parentheses). $\psi_0$: insertion phase in the DE405 body frame (Section~\ref{sec:sim_frame}). $\bar e_{\min}$: smallest slow eccentricity (27.3-day running mean of the eccentricity vector, measured from the lunar equator) and the day it occurs; $\Delta\bar\omega$: net rotation of the slow eccentricity vector over the mission; cells: latitude--altitude cells estimated as crossed (per-revolution count), of 330.}
\label{tab:sim_scan}
\setlength{\tabcolsep}{2pt}
\begin{tabular}{rcccccc}
\toprule
$\Omega_0$ (deg) & $\psi_0$ (deg) & Lowest (km) & Highest (km) & $\bar e_{\min}$ (day) & $\Delta\bar\omega$ (deg) & Cells \\
\midrule
0 & 85 & 37.8 (39.3) & 222.8 & 0.0021 (459) & $-375.7$ & 261 (317) \\
30 & 115 & 39.9 (42.0) & 220.7 & 0.0001 (401) & $-10.8$ & 252 (309) \\
60 & 145 & 31.6 (35.0) & 229.2 & 0.0020 (389) & $-365.6$ & 261 (308) \\
90 & 175 & 36.5 (41.8) & 224.3 & 0.0005 (389) & $-365.7$ & 254 (293) \\
120 & 205 & 41.4 (49.6) & 219.1 & 0.0016 (388) & $-4.8$ & 245 (281) \\
150 & 235 & 55.3 (55.1) & 204.7 & 0.0064 (434) & $-9.1$ & 219 (275) \\
180 & 265 & 55.3 (55.2) & 204.6 & 0.0068 (490) & $-8.4$ & 213 (277) \\
210 & 295 & 53.1 (53.3) & 207.0 & 0.0063 (519) & $-3.9$ & 222 (282) \\
240 & 325 & 58.8 (58.9) & 201.7 & 0.0096 (539) & $+1.9$ & 200 (256) \\
270 & 355 & 46.8 (47.6) & 213.8 & 0.0035 (568) & $-0.8$ & 237 (293) \\
300 & 25 & 33.9 (36.2) & 226.9 & 0.0017 (604) & $-360.2$ & 261 (312) \\
330 & 55 & 35.4 (36.2) & 225.2 & 0.0018 (540) & $-369.5$ & 261 (317) \\
\bottomrule
\end{tabular}
\end{table}

\begin{figure}[htbp]
\centering
\includegraphics[width=0.98\textwidth]{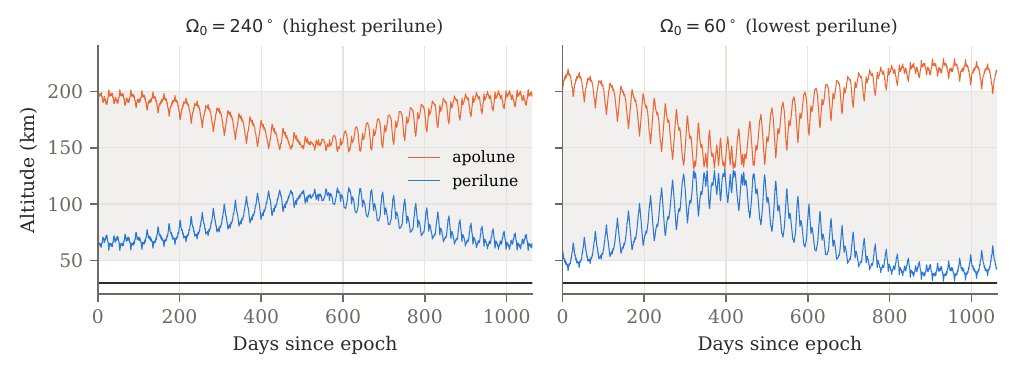}
\caption{Perilune and apolune altitude at every revolution ($50\times50$ field) for the highest-perilune case ($\Omega_0=240^\circ$, left) and the lowest ($\Omega_0=60^\circ$, right). Shaded: science shell; line: 30~km floor.}
\label{fig:alt_hist}
\end{figure}

\subsection{Circulation of the mean periapsis and coverage}
\label{sec:sim_circ}
\begin{table}[htbp]
\centering
\small
\caption{Net rotation $\Delta\bar\omega$ (deg) of the slow eccentricity vector of the circulating node-scan runs: with the nominal 27.3-day running mean over days 14--1049, and with six filters over the common span of days 27.5--1035 (``27+13.5'': a 27.0-day mean followed by a 13.5-day mean).}
\label{tab:r4_filter}
\setlength{\tabcolsep}{3pt}
\begin{tabular}{lccccccc}
\toprule
 & Nominal & \multicolumn{6}{c}{Common span, filter (days)} \\
Run & 27.3, 14--1049 & 27.3 & 27.0 & 25 & 30 & 27+13.5 & 54 \\
\midrule
T000 & $-375.7$ & $-370.4$ & $-370.6$ & $-371.8$ & $-369.0$ & $-370.6$ & $-370.6$ \\
T060 & $-365.6$ & $-360.5$ & $-360.6$ & $-361.7$ & $-359.3$ & $-360.6$ & $-360.7$ \\
T090 & $-365.7$ & $-360.5$ & $-360.7$ & $-361.9$ & $-359.2$ & $-360.7$ & $-360.7$ \\
T300 & $-360.2$ & $-355.9$ & $-355.8$ & $-354.7$ & $-357.1$ & $-355.8$ & $-355.8$ \\
T330 & $-369.5$ & $-364.7$ & $-364.7$ & $-365.1$ & $-364.2$ & $-364.7$ & $-364.7$ \\
\bottomrule
\end{tabular}
\end{table}

In what follows, singly averaged elements keep the monthly terms; the slow (monthly-filtered) eccentricity vector $\bar\varepsilon$ removes them: it is obtained by applying the same 27.3-day running mean to the complete output of GMAT and of the full single-averaged model, so that the two are compared with the same filter (the $\bar\varepsilon$ of Eq.~\eqref{eq:forced_sol} is used only in the two-step prediction); ``mean eccentricity'' in Tables~\ref{tab:sim_scan} and~\ref{tab:sim_target} refers to it. In every run of the scan the slow eccentricity becomes small in the middle of the mission: its minimum, 0.0001--0.0096, occurs between days 388 and 604 (Table~\ref{tab:sim_scan}), and the propagated histories show that the osculating eccentricity, the sum of the slow and the monthly forced vectors, then oscillates within each month between nearly zero and about 0.015, so that the orbit is close to circular at about 130~km for part of each month (Fig.~\ref{fig:alt_hist}). Whether the mean periapsis completes a circulation depends on the side on which the slow eccentricity vector passes the origin. It does at five nodes ($\Delta\bar\omega=-360^\circ$ to $-376^\circ$); at the other seven the vector passes the origin on the other side and the net rotation is between $-11^\circ$ and $+2^\circ$, so that the mean periapsis returns instead of circulating and the circulation required by R4 is not completed. With the $20\times20$ field the same initial states give a smallest slow eccentricity of 0.004--0.019 and a circulation of the slow vector at all twelve nodes.

\begin{figure}[htbp]
\centering
\includegraphics[width=0.98\textwidth]{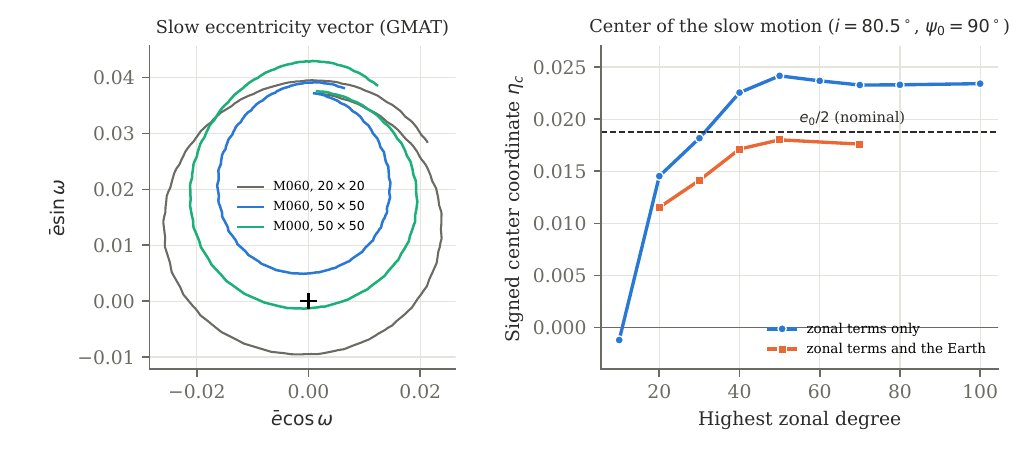}
\caption{Left: slow eccentricity vector (27.3-day running mean, measured from the lunar equator) of the targeted runs M060 with the $20\times20$ and $50\times50$ fields and M000 with $50\times50$; $+$: origin. Right: signed coordinate $\eta_c$ of the frozen center of the slow motion (equal to $e_c$ where positive for the zonal terms only, which put the center on the axis $\omega=90^\circ$; approximately equal with the Earth, which moves it slightly off this axis) versus the highest zonal degree ($a=1868$~km, $i=80.5^\circ$ about the lunar equator, two-dimensional equilibrium), with the zonal terms only and with the Earth added (monthly average, fixed node $\psi_0=90^\circ$, Section~\ref{sec:sim_circ}), and half the nominal eccentricity (the circular-contour threshold $e_0=2e_c$; the origin-crossing threshold at $\omega_0=90^\circ$ of the fixed-$i,\Omega$ reduced model is $e_s=0.0374$--0.0383 depending on the node, Section~\ref{sec:sim_circ}).}
\label{fig:slow_centre}
\end{figure}

The cause is the location of the center of the slow motion (Fig.~\ref{fig:slow_centre}). Under the zonal terms and the Earth the slow eccentricity vector rotates about a frozen center $\varepsilon_c=\xi_c+\mathrm i\eta_c$ on (with the Earth, close to) the axis $\xi=0$. We write $e_c=|\varepsilon_c|\ge0$ for its distance from the origin and $\eta_c$ for its signed coordinate, positive when the center lies on the side $\omega=90^\circ$, as for the nominal orbit ($\eta_c\approx e_c$). For a start at $\omega_0=90^\circ$ with $\eta_c>0$ and eccentricity $e_0>e_c$ the rotation has radius $e_0-e_c$, so that, if the slow contour were a circle about the center, the vector would pass the origin at a distance of about $|e_0-2e_c|$ and enclose it only if $e_0>2e_c$. The contours are not exactly circles. In these reduced analyses the fixed node is measured in the equator frame of the epoch from the prime meridian at the epoch, $\Omega'=\Omega-\theta_0$, which equals the insertion phase $\psi_0$ of Eq.~\eqref{eq:savg_psi}; we therefore label it by $\psi_0$. All reduced-model values in this section, in Fig.~\ref{fig:slow_centre} and in Section~\ref{sec:design} use this convention. With the zonal terms to degree~50 and the Earth averaged over a month, $e_c$ at $80.5^\circ$ is 0.01843 at $\psi_0=0^\circ$, 0.01824 at $45^\circ$, 0.01802 at $90^\circ$ and 0.01798 at $135^\circ$, so the circular estimate is $2e_c=0.0360$--0.0369, slightly below the nominal $e_0=0.03747$. We define $e_s$ as the origin-crossing threshold of a reduced model: the month-averaged single-averaged nonsingular equations (Earth sampled at eight equally spaced times of the month starting at the epoch; the same sample set is used at every evaluation, so the reduced vector field is autonomous and is held fixed throughout each reduced-model integration and in the equilibrium computations) are propagated with a step of 1~day from $(\xi,\eta)=(10^{-6},0)$ with $a$, $i$ and $\Omega$ held fixed, in the equator frame of the epoch, and $e_s$ is the eccentricity at the first nontrivial crossing of the positive $\eta$ axis ($\omega=90^\circ$) after departure from the near-origin initial state, located by linear interpolation between steps. Starting from $(10^{-5},0)$ or $(0,10^{-6})$ gives the same $e_s$ to $10^{-5}$, and halving the step to 0.5~day changes $e_s$ at $\psi_0=90^\circ$ by less than $10^{-5}$. Because the month-averaged Earth term depends on the node, so does $e_s$: it is 0.0383 at $\psi_0=0^\circ$, 0.0379 at $45^\circ$, 0.0375 at $90^\circ$, 0.0374 at $135^\circ$ and 0.0377 at $180^\circ$. Doubling the number of Earth samples to 16 changes $e_s$ by less than $10^{-5}$ at $\psi_0=0^\circ$ and $90^\circ$, and 16 or 32 samples leave $e_c$ at $80.5^\circ$ and the inclinations at which $\eta_c=0$ unchanged to $10^{-5}$ and $10^{-3}$~deg. Holding $\mathcal H=\mathcal G\cos i$ fixed instead of $i$ changes $e_s$ at $\psi_0=90^\circ$ by less than $10^{-5}$, so the reduction itself is not the cause of the spread, and an independent doubly averaged calculation with the same field (0.0377) lies within it. The nominal $e_0=0.0375$ lies inside this range, which is consistent with the dependence of the circulation on the node in the GMAT runs; since the node regresses by about $170^\circ$ during the mission, no single fixed-$\Omega$ value is the threshold of the full problem. For a fixed contour of the reduced model and a start at $\omega_0=90^\circ$ the condition is $e_0>e_s$, and the circular estimate is an approximation of it. The largest eccentricity of a contour (0.0385 near $\omega=86^\circ$ at $\psi_0=0^\circ$) is a different quantity and is not the threshold for a given initial angle. At $i=80.5^\circ$ about the lunar equator, with the zonal terms alone, $e_c$ is 0.015 with $J_2$--$J_{20}$, 0.018 with $J_2$--$J_{30}$ and 0.023--0.024 from degree~50 to degree~100. The Earth, averaged over a month, moves the center toward the origin: $e_c$ is 0.012 with $J_2$--$J_{20}$, 0.014 with $J_2$--$J_{30}$, 0.018 with $J_2$--$J_{50}$ and 0.0176 with $J_2$--$J_{70}$ at $\psi_0=90^\circ$, and varies by about $4.5\times10^{-4}$ over the sampled nodes; with the Earth the center lies close to, not exactly on, the axis $\omega=90^\circ$ ($\xi_c\approx0.02\,\eta_c$). The nominal $e_0=0.0375$ lies just above the circular estimate; on the fixed contour at $\psi_0=0^\circ$ ($e_s=0.0383$) the nominal start does not enclose the origin, and its smallest slow eccentricity is only 0.0006 (day 509). Thus the mean vector passes very close to the origin and the side on which it passes depends on the node, on the initial inclination and on the field beyond degree~20; with $J_2$--$J_{20}$ and the Earth, $e_0-2e_c=0.014$ and the vector encloses the origin with a wide margin, which is what the $20\times20$ runs show. When the slow eccentricity is small, the monthly forced term, $|\delta\varepsilon|\le0.014$, dominates; the osculating eccentricity is then bounded by about 0.014 plus the slow eccentricity, and in the propagated histories it comes close to zero during part of each month, although the inequality alone does not guarantee this. We classify a run as circulating if the net rotation $\Delta\bar\omega$ of the slow vector, unwrapped over the mission after a 27.3-day running mean, exceeds $300^\circ$ in magnitude; the runs fall into two groups, $|\Delta\bar\omega|=360$--$376^\circ$ (T300: $-360.2^\circ$) and $|\Delta\bar\omega|\le11^\circ$, with no intermediate case. This classification describes the topology of the slow motion; requirement R4, at least one full circulation within the mission, is judged separately by $|\Delta\bar\omega|\ge360^\circ$ over the span of the running mean (days 14--1049). Under this window and span, five of the twelve node-scan runs (six of the twenty runs) satisfy R4, but three of them, T060, T090 and T300, exceed one revolution by only $5.6^\circ$, $5.7^\circ$ and $0.2^\circ$. To separate the filter from the span, we repeated the analysis with running means of 25, 27.0, 27.3, 30 and 54~days and with a cascaded 27.0-day and 13.5-day mean, which tests the sensitivity to the filter response near the first two forcing harmonics, all evaluated over the common span of days 27.5--1035. The net rotation of T060 and T090 is then $-359.2^\circ$ to $-361.9^\circ$ and that of T300 $-354.7^\circ$ to $-357.1^\circ$. Five of the twelve runs meet R4 under the nominal filter and span, and two to four meet it over the shorter common span, depending on the filter: T300 falls short of one revolution over the common span with every filter, so its status depends mainly on the span, whereas T060 and T090 cross $360^\circ$ there depending on the filter. The classification ($|\Delta\bar\omega|>300^\circ$) is the same in all twenty runs for all filters, including the near-origin runs T030 and T090, and the smallest slow eccentricity changes by at most 0.0004. Completion of one revolution is therefore marginal for several circulating cases (Table~\ref{tab:r4_filter}); a net rotation below $360^\circ$ over the shortened common span does not by itself mean that the full mission falls short of one revolution; the per-run values are given in Tables~\ref{tab:sim_scan} and~\ref{tab:sim_target}. The nonsingular single-averaged model gives the same classification as GMAT in 19 of the 20 runs (the exception, T090, has $\bar e_{\min}=0.0005$) and its smallest slow eccentricity within 0.001 of GMAT. With degree~70 and~100 the model gives a circulation for T090; the classification is not robust when the vector passes this close.

\begin{figure}[htbp]
\centering
\includegraphics[width=0.98\textwidth]{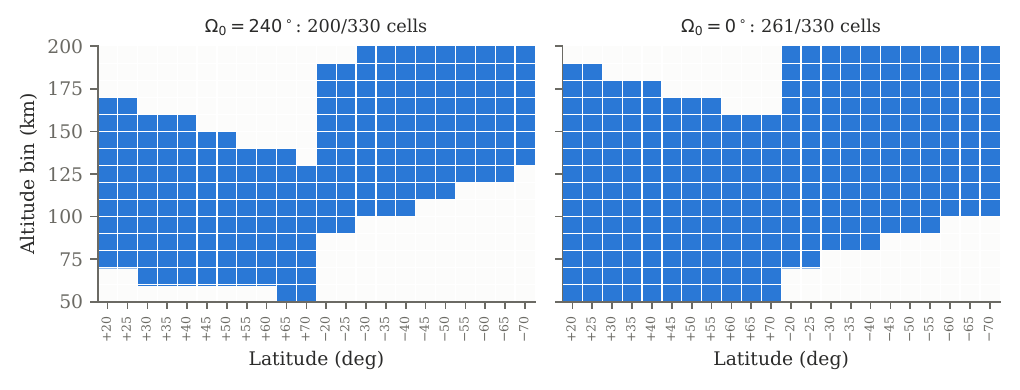}
\caption{Latitude--altitude cells estimated as crossed at least once in three lunar years ($50\times50$ field; filled; per-revolution count) for the case with the fewest cells ($\Omega_0=240^\circ$, left) and one with the most ($\Omega_0=0^\circ$, right).}
\label{fig:coverage}
\end{figure}

The near-circular phase reduces the coverage. Between 200 and 261 of the 330 cells are estimated to be crossed (61--79\%; altitudes for the cells, R2 and R3 are measured from 1738.2~km, as defined in Section~\ref{sec:req}), against 256--317 (78--96\%) with the $20\times20$ field; no run crosses all 330 cells, so R1 is not satisfied by any tested case. While the orbit is nearly circular, neither the 50--60~km nor the 190--200~km bins are visited, and the cells missed lie at the bottom and the top of the shell (Fig.~\ref{fig:coverage}). The runs whose mean periapsis circulates cross 254--261 cells, the others 200--252.

Cells are counted per revolution: with the osculating elements at each perilune passage, measured from the lunar equator, the altitudes of the two crossings of each signed latitude ($u_1$ and $\pi-u_1$ for $+\varphi$, $\pi+u_1$ and $2\pi-u_1$ for $-\varphi$) are evaluated with Eq.~\eqref{eq:thy_h}, and a cell counts as crossed if any revolution places a crossing in it. This neglects the change of the elements within one revolution, so the counts are estimates; they have not yet been checked against latitude crossings detected in densely sampled states. Because $e$ changes along the circulation, the lowest crossing of a latitude need not occur when the perilune is above that latitude; the per-revolution evaluation does not assume that it does.

The condition $e_0>2e_c$ is the geometric condition for a circular slow contour started at $\omega_0=90^\circ$ to enclose the origin, and $e_0>e_s$ the corresponding condition for a fixed contour of the averaged model and a start at $\omega_0=90^\circ$; for a moving center or a non-circular contour the actual time history decides, and neither is a general necessary condition. Both thresholds refer to the slow (mean) eccentricity, whereas $e_0$ of the untargeted runs is the osculating value; the mean initial vector $\bar\varepsilon_0=\varepsilon_0-\delta\varepsilon(\psi_0)$ differs from $\varepsilon_0$ by up to $|\delta\varepsilon|_{\max}\approx0.014$, an order of magnitude more than the spread of $e_s$ over the node (about 0.001). The comparison of $e_0$ with $2e_c$ and $e_s$ therefore shows that the nominal orbit lies close to the transition, but it does not predict which individual runs circulate; that classification is taken from the propagations. A sufficient, not necessary, margin against passages near zero is that the smallest slow eccentricity along the trajectory, $\bar e_{\min}$, exceeds the forced amplitude $|\delta\varepsilon|_{\max}$; then the osculating eccentricity remains bounded away from zero by $M_e$, within the decomposition. Since $|\varepsilon(t)|\ge|\bar\varepsilon(t)|-|\delta\varepsilon(t)|$, this margin, $M_e=\min_t|\bar\varepsilon|-\max_t|\delta\varepsilon|>0$, excludes near-circular passages but does not imply circulation: a slow trajectory that returns without enclosing the origin can have a positive margin as well. The circulation is therefore checked separately, from the net rotation of the slow phase over the mission. We evaluate $\bar e_{\min}$ from the propagated slow motion and do not replace it by its circular approximation $e_0-2e_c$, which is of the same size as the margin itself for the nominal orbit. On the fixed slow contour at $\psi_0=0^\circ$, $\bar e_{\min}$ is 0.0006 for the nominal $60\times200$~km orbit, against $|\delta\varepsilon|_{\max}\approx0.014$. The remedies are a larger eccentricity or an inclination at which the frozen center lies closer to the origin (Section~\ref{sec:design}).

\subsection{Mean-eccentricity targeting}
\label{sec:sim_target}
To test the design rule~\eqref{eq:forced_target}, eight runs start from osculating elements chosen so that the mean eccentricity vector equals the design value at the insertion phase: six runs of the nominal $60\times200$~km orbit at $\Omega_0=0^\circ,60^\circ,\dots,300^\circ$ (M000--M300), and two runs with a 70~km mean perilune at the two worst nodes of the untargeted scan, $60^\circ$ and $300^\circ$ (H060, H300). Table~\ref{tab:sim_target} gives the osculating eccentricity and the rotation of the line of apsides required by Eq.~\eqref{eq:forced_target}, and the lowest altitude predicted by Eq.~\eqref{eq:forced_twostep} and obtained with GMAT, over the first 300 days and over three lunar years. In all eight runs the GMAT initial state agrees with the intended state to $10^{-10}$~km; this is the numerical consistency of the full-precision input, not an accuracy of the orbit, and the states listed in \ref{app:states} are rounded. All eight runs cover the full three lunar years.

\begin{table}[htbp]
\centering
\small
\caption{Mean-eccentricity targeting ($50\times50$ field): required osculating eccentricity $e_0$ and apsidal rotation $\Delta\omega_0$, lowest altitude over the first 300 days predicted by Eq.~\eqref{eq:forced_twostep} and over three lunar years given by the full single-averaged model, and the GMAT values. The untargeted GMAT values over three lunar years at the same nodes are 37.8, 31.6, 41.4, 55.3, 58.8 and 33.9~km (Table~\ref{tab:sim_scan}). Cells: latitude--altitude cells estimated as crossed, of 330. $\Delta\bar\omega$: net rotation of the slow eccentricity vector (GMAT).}
\label{tab:sim_target}
\setlength{\tabcolsep}{2pt}
\begin{tabular}{rccccccccc}
\toprule
$\Omega_0$ & Mean orbit & $e_0$ & $\Delta\omega_0$ & \multicolumn{2}{c}{Days 0--300 (km)} & \multicolumn{2}{c}{Three years (km)} & Cells & $\Delta\bar\omega$ \\
(deg) & (km) & & (deg) & Eq.~\eqref{eq:forced_twostep} & GMAT & model & GMAT & & (deg) \\
\midrule
0 & 60$\times$200 & 0.0365 & $-3.1$ & 50.1 & 50.2 & 40.1 & 39.9 & 255 & $-375.3$ \\
60 & 60$\times$200 & 0.0355 & $+18.0$ & 50.8 & 50.7 & 46.9 & 46.8 & 227 & $-7.8$ \\
120 & 60$\times$200 & 0.0398 & $+12.7$ & 51.1 & 50.7 & 46.1 & 46.2 & 234 & $-6.2$ \\
180 & 60$\times$200 & 0.0399 & $-2.4$ & 51.2 & 50.9 & 51.4 & 50.9 & 227 & $-9.1$ \\
240 & 60$\times$200 & 0.0422 & $-4.5$ & 50.7 & 50.9 & 51.0 & 50.9 & 225 & $+1.9$ \\
300 & 60$\times$200 & 0.0355 & $-20.6$ & 49.8 & 50.3 & 49.8 & 49.8 & 227 & $-0.9$ \\
60 & 70$\times$200 & 0.0328 & $+19.1$ & 60.9 & 60.9 & 57.4 & 57.2 & 222 & $-10.1$ \\
300 & 70$\times$200 & 0.0329 & $-21.8$ & 59.9 & 60.5 & 59.9 & 60.1 & 214 & $-3.4$ \\
\bottomrule
\end{tabular}
\end{table}

Over the first 300 days the six nominal runs reach 50.2--50.9~km, against 33.9--58.9~km for the untargeted runs at the same nodes over the same period, and Eq.~\eqref{eq:forced_twostep} predicts all eight targeted runs within 0.6~km. In this phase the spread over the node is reduced from 25.0~km to 0.7~km, as designed. Over three lunar years the spread is 11.0~km (39.9--50.9~km) instead of 27.2~km, and the worst case rises from 31.6~km to 39.9~km. The remaining spread develops in the second half of the mission through the slow motion; the full single-averaged model reproduces the mission-minimum perilune altitude of these runs within 0.5~km. With a 70~km mean perilune the lowest altitude at the two worst nodes is 57.2 and 60.1~km, 27--30~km above the floor. Targeting does not restore the circulation: it sets the mean eccentricity to the design value, which still lies close to $2e_c$ and within the node-dependent range of $e_s$, and the mean periapsis circulates only in M000. Matched by node, targeting changes the number of cells by $-6$, $-34$, $-11$, $+14$, $+25$ and $-34$ at $\Omega_0=0^\circ,60^\circ,\dots,300^\circ$ (225--255 cells against 200--261 without targeting): it raises the coverage where the untargeted perilune stays high and lowers it where it removes the deepest perilune passes, so it improves the worst-case perilune without improving the coverage.

\subsection{First-order inclination law}
\label{sec:sim_icorr}
Runs I5--I8 (intermediate set, $20\times20$ field) repeat the nodes $0^\circ$, $90^\circ$, $180^\circ$ and $270^\circ$ with $i_0=80.5^\circ+0.606^\circ\cos2\psi_0$. The lowest altitudes are 53.3, 35.8, 41.6 and 53.2~km, against 53.3, 36.0, 41.4 and 53.0~km for the same nodes and initial state without the correction. The correction leaves the perilune unchanged within 0.2~km, consistent, for these $20\times20$ cases and this range of inclination correction, with the finding that the perilune margin is controlled by the forced eccentricity rather than by the inclination; the near-critical slow motion with the $50\times50$ field was not tested in this way.

\section{Implications for orbit design}
\label{sec:design}
The mechanisms of Sections~\ref{sec:forced} and~\ref{sec:sim} reduce to three checks for a maintenance-free orbit that must sample an altitude shell. All three can be evaluated with the nonsingular single-averaged model of Section~\ref{sec:savg}, without full-force propagation.

\paragraph{Circulation} The slow phase of the mean eccentricity vector must complete the required rotation over the mission. Three questions are involved: whether the contour of the fixed-node reduced model encloses the origin, whether the actual slow phase circulates once the center and the orbit plane evolve, and whether one revolution is completed within the mission. In the fixed-node reduced model, the origin-crossing threshold $e_s$ (Section~\ref{sec:sim_circ}), of which $2e_c$ is a circular approximation, provides a diagnostic for the first, for a start at $\omega_0=90^\circ$ with the center on the same side ($\eta_c>0$). For the evolving-node problem, the circulation and its completion within the mission must be checked from the propagated slow phase. The nominal eccentricity lies within the node-dependent range of $e_s$, and only 6 of the 20 runs belong to the circulating group.

\paragraph{Margin against near-circular passages} The margin $M_e=\bar e_{\min}-|\delta\varepsilon|_{\max}$ should be positive, so that the osculating eccentricity remains bounded away from zero by $M_e$; an engineering application would specify a minimum acceptable $M_e$. It does not imply circulation (Section~\ref{sec:sim_circ}). For the nominal orbit $\bar e_{\min}$ is 0.0001--0.0096 in the GMAT runs and 0.0006 on the fixed contour of the averaged model, against $|\delta\varepsilon|_{\max}\approx0.014$, so the margin is negative at every node.

\paragraph{Perilune screen for a targeted orbit} After targeting the mean eccentricity, Eq.~\eqref{eq:forced_target}, the lowest perilune is estimated by
\begin{equation}
\hat h_{p,\min}=\bar h_p-A-\Delta h_{\rm slow},
\label{eq:design_margin}
\end{equation}
and the screen requires $\hat h_{p,\min}\ge h_{\rm floor}$, that is, $\bar h_p\ge h_{\rm floor}+A+\Delta h_{\rm slow}$. Here $\bar h_p$ is the mean perilune altitude, $A=a|\delta\varepsilon|_{\max}$ the forced amplitude and $\Delta h_{\rm slow}$ the decrease of the perilune caused by the slow motion. For the targeted $60\times200$~km orbit, $\bar h_p=59.8$~km, $A=25.1$~km (Table~\ref{tab:forced_amp}, degree~50) and $\Delta h_{\rm slow}=0.1$~km from the single-averaged model at $\psi_0=90^\circ$, so Eq.~\eqref{eq:design_margin} gives 34.6~km (altitudes above 1738.2~km); for the targeted $70\times200$~km orbit, with $A=24.7$~km, it gives 45.0~km. GMAT with the $50\times50$ field gives 39.9--50.9~km for the six targeted nominal runs and 57.2 and 60.1~km for H060 and H300 (Table~\ref{tab:sim_target}): the estimate lies below the observed minima in all eight tested cases, by 5--16~km, consistent with incomplete alignment of the slow and forced vectors; differences in the slow evolution and the decomposition residual also contribute. Equation~\eqref{eq:design_margin} is a screen, not a strict bound: $A$ is evaluated at fixed elements, $\Delta h_{\rm slow}$ for one initial node, and the residual of the decomposition~\eqref{eq:forced_sol} is not included. It applies only to a targeted orbit, because without targeting the mean eccentricity differs from the design value by up to $|\delta\varepsilon|_{\max}$ (Section~\ref{sec:forced_node}).

\paragraph{Two controls of the circulation} The first two checks can be met by raising the eccentricity, so that $e_0$ exceeds the threshold with a margin, or by moving the frozen center. The center depends on the inclination. We solve the two-dimensional equilibrium $\dot{\bar\varepsilon}=0$ of the month-averaged slow equations (zonal terms to degree~50 and the Earth, $a=1868$~km, fixed node $\psi_0=90^\circ$ in the convention of Section~\ref{sec:sim_circ}) for $(\xi_c,\eta_c)$. The residual $|\dot{\bar\varepsilon}(\varepsilon_c)|$ is below $10^{-16}$~day$^{-1}$, i.e.\ below $10^{-12}$ of the characteristic slow rate $|\dot{\bar\varepsilon}|\approx1.1\times10^{-4}$~day$^{-1}$ evaluated at $\varepsilon=0.0375\,\mathrm i$. The signed coordinate $\eta_c$ is $+0.018$ at $80.5^\circ$, $+0.0036$ at $77^\circ$ and $+0.0030$ at $84^\circ$, and $-0.0016$ at $76^\circ$ and $-0.0038$ at $85^\circ$, while $\xi_c\approx0.02\,\eta_c$, so the center moves along a path that is nearly a straight line through the origin. Solving $\eta_c(i)=0$ to $10^{-6}$~deg gives $i=76.33^\circ$ and $84.46^\circ$, where $|\eta_c|<10^{-10}$ and $|\xi_c|=7\times10^{-8}$ and $7\times10^{-7}$: the frozen center approaches the origin closely, where its signed coordinate changes sign. We do not claim an exact crossing, which would require both components to vanish simultaneously. The linearized equilibrium is elliptic: the eigenvalues of the Jacobian $\partial(\dot\xi,\dot\eta)/\partial(\xi,\eta)$ at $(\xi_c,\eta_c)$ are numerically purely imaginary, with $|\mathrm{Re}/\mathrm{Im}|<4\times10^{-6}$ at $76^\circ$, $77^\circ$, $80.5^\circ$, $84^\circ$ and $85^\circ$ (unchanged for difference steps of $10^{-5}$ to $10^{-7}$), and small-amplitude periods of 1045, 1029, 967, 760 and 754~days. This linear analysis does not exclude a very weak nonlinear focus. At $80.5^\circ$ one revolution near the center therefore takes about 970~days, comparable to the 1063-day mission, which is consistent with the marginal completion of one revolution in Table~\ref{tab:r4_filter}. This is a local timescale comparison, not a prediction of the finite-amplitude circulation period. With GRGM1200A \cite{Goossens2016,Goossens2020} truncated to degree~50 (reference radius 1738.0~km, fully normalized coefficients in the principal-axis frame, as for LP165P; the GM of LP165P is used, which differs by $9.3\times10^{-4}$~km$^3$/s$^2$, a relative difference of $2\times10^{-7}$) the zero crossings of $\eta_c$ occur at $76.42^\circ$ and $84.49^\circ$, and at $80.5^\circ$ $\eta_c=0.018016$ for LP165P and 0.018081 for GRGM1200A, a difference of $6.5\times10^{-5}$ (Table~\ref{tab:centre_fields}).

\begin{table}[htbp]
\centering
\small
\caption{Frozen center of the month-averaged slow problem (zonal terms to degree~50 and the Earth, $a=1868$~km, fixed node $\psi_0=90^\circ$, Section~\ref{sec:sim_circ}) for two degree-50 fields: signed coordinate $\eta_c$ and inclinations at which $\eta_c=0$.}
\label{tab:centre_fields}
\setlength{\tabcolsep}{3pt}
\begin{tabular}{lccccccc}
\toprule
Field & \multicolumn{5}{c}{$\eta_c$ at $i=$} & \multicolumn{2}{c}{$\eta_c=0$ at $i=$} \\
 & $76^\circ$ & $77^\circ$ & $80.5^\circ$ & $84^\circ$ & $85^\circ$ & first & second \\
\midrule
LP165P & $-0.0016$ & $+0.0036$ & $+0.01802$ & $+0.0030$ & $-0.0038$ & $76.33^\circ$ & $84.46^\circ$ \\
GRGM1200A & $-0.0021$ & $+0.0033$ & $+0.01808$ & $+0.0032$ & $-0.0035$ & $76.42^\circ$ & $84.49^\circ$ \\
\bottomrule
\end{tabular}
\end{table}

These are equilibria of a fixed-node reduced model; the node dependence of the Earth term and the regression of the node during the mission are not included. Where the center is close to the origin, the condition $e_0>2e_c$, derived for a center on the side of the initial periapsis, does not carry over unchanged, and a small eccentricity may suffice for circulation. Which of the two controls is preferable depends on the shell to be sampled, the safety floor and the latitude band; a design method built on these checks, with full-force verification, is the subject of a companion paper.

\section{Discussion}
\label{sec:discussion}

\paragraph{Which model to design with}
The doubly averaged model gives the drift rate and the structure of the slow motion. It does not give the perilune margin, and with a truncated zonal series it misplaces the center of the slow motion. Two corrections are needed, and both are cheap: the zonal series must be carried to degree $\approx50$ (degree $\approx30$ suffices for the lowest perilune of the zonal problem, Table~\ref{tab:savg_conv}, but not for the frozen center, Fig.~\ref{fig:slow_centre}), and the forced eccentricity of Eq.~\eqref{eq:forced_sol} must be added. Both are contained in the nonsingular single-averaged model, which costs a few minutes per three-year run with the $50\times50$ field and reproduces the mission-minimum perilune altitude of all twenty GMAT runs within 0.5~km, as well as the smallest slow eccentricity and, in 19 of the 20 runs, the sense of the slow rotation; the two-step prediction of Eq.~\eqref{eq:forced_twostep} is within 1.2~km for horizons up to 700 days and overestimates the lowest perilune by up to 9.3~km at intermediate horizons (7.7~km at the mission end).

\paragraph{Why low lunar orbits are sensitive to the tesseral field}
The forced eccentricity is large for three reasons that are particular to the Moon: the $l=3$, $m=1$ coefficients are larger than $C_{22}$; the slow rotation multiplies their effect by $n/\dot\theta\approx326$; and the forcing acts on $\varepsilon$ directly, independently of $e$ to leading order, so that it matters most for the small eccentricities of low orbits. For near-circular and frozen orbits this is the translation theorem exploited for station keeping \cite{Holt2024,Yarndley2025}. For a maintenance-free circulating orbit the same term has a different role: it cannot be corrected, so it sets the perilune margin, and through the insertion phase it sets the node dependence.

\paragraph{Selecting the insertion}
The lowest perilune depends on the initial node through $\delta\varepsilon(\psi_0)$, i.e.\ through the phase of the Moon's rotation at insertion. Two strategies follow. The node and insertion time can be chosen so that $\delta\varepsilon(\psi_0)$ points along the initial eccentricity vector, which lowers the mean eccentricity and gives the best cases of Table~\ref{tab:sim_scan}. Alternatively, the insertion can target the mean eccentricity, Eq.~\eqref{eq:forced_target}, which reduces the node dependence to about 1~km in the first 300 days and, within the first-order model and as long as the slow motion does not raise $|\bar\varepsilon|$ above its design value, gives the conditional bound $a(1-|\bar\varepsilon^\ast|-|\delta\varepsilon|_{\max})-R_{\rm alt}$; the full-force runs show that a smaller node dependence returns later in the mission. In both cases the node is a design variable of the transfer, not a free parameter of the science orbit.

\paragraph{Margin versus coverage}
With the nominal perilune of 60~km and the $50\times50$ field, the worst case of the scan keeps 1.6~km above the 30~km floor (T060), and in the model the field to degree~70 raises it by 0.8~km. Restoring a margin for insertion errors means raising the mean perilune, at the cost of the lowest bins of the shell. Targeting the mean eccentricity is the first remedy: it raises the worst case of the nominal design from 31.6~km to 39.9~km (Section~\ref{sec:sim_target}). Raising the mean perilune to 70~km is the second; with targeting the lowest altitude at the worst nodes is 57.2--60.1~km. Neither restores the coverage, which is limited by the near-circular phase of Section~\ref{sec:sim_circ}. Circulation provides the intended latitude--altitude sweep, but the coverage must be evaluated directly; a margin against near-circular passages requires the slow eccentricity to stay clear of zero, $\bar e_{\min}>|\delta\varepsilon|_{\max}$; the nominal orbit meets the first only at some nodes and the second at none. Section~\ref{sec:design} shows that both can be addressed through the eccentricity or through the inclination, which sets the position of the frozen center.

\paragraph{Limitations}
The single-averaged model is first order in the gravity field and neglects short-period terms; with the rotation model of the propagator it reproduces the lowest altitude of the decomposition cases of Table~\ref{tab:savg_valid} within 0.3~km and the mission-minimum perilune altitude of the twenty main runs within 0.5~km. The bound~\eqref{eq:forced_bound} is attained only if the forced and slow eccentricity vectors align, so it overestimates the decrease when many frequencies combine (Table~\ref{tab:forced_amp}); it also holds only within the first-order decomposition. The two-step prediction separates slow and forced motion; when the slow eccentricity passes close to zero this separation fails, and the prediction overestimates the late perilune by up to 9.3~km (7.7~km at the mission end), for reasons only partly identified (Section~\ref{sec:forced_node}); for the tested cases, kilometer-level accuracy is demonstrated only up to approximately 700 days. Model and propagator must use the same truncation of the field, and the truncation must reach degree $\approx50$ for the slow motion; the main GMAT runs use a $50\times50$ field, which the model shows to be conservative for the lowest perilune by up to 2~km at the nodes evaluated at higher degree (an extrapolation for the others) and, in the case tested (T090), to leave the cell count within 4 cells of the counts at degree~70 and~100. Where the slow eccentricity vector passes within about 0.001 of the origin, the sense of its rotation, and therefore whether the mean periapsis circulates, is not robust to the field truncation (T090; only this node was tested at degree~70 and~100). The nominal eccentricity is close to the circulation threshold of the reduced model (0.0374--0.0383), which refers to the mean eccentricity and serves as a diagnostic, not a per-run prediction; all propagations use LP165P. The tested frozen-center locations are similar in LP165P and GRGM1200A at degree~50 (Table~\ref{tab:centre_fields}); however, the sensitivity of the circulation count and of the finite-mission extrema to the gravity model has not been evaluated, and the coefficient uncertainties have not been propagated. The frozen-center values of Section~\ref{sec:design} are computed for one fixed node, $\psi_0=90^\circ$ (Section~\ref{sec:sim_circ}). The cell counts are estimated per revolution from the osculating elements and have not been compared with latitude crossings detected directly in the propagated trajectory, and the longitude coverage has not been verified, so the percentages are not a three-dimensional coverage. Altitudes are above a reference sphere; terrain clearance and insertion errors are not included in R2.

\section{Conclusions}
\label{sec:concl}

A low eccentric lunar orbit left without maneuvers can sample a large part of the 50--200~km shell between $20^\circ$ and $70^\circ$ latitude, but its perilune margin and its coverage are controlled by two different features of the lunar field. For the $60\times200$~km design at $80.5^\circ$:
\begin{enumerate}
\item Twenty full-force propagations over three lunar years with a $50\times50$ field show no impact; the lowest altitude is 31.6--58.8~km over twelve nodes, and an estimated 61--79\% of the shell cells are crossed.
\item The perilune margin is set by the monthly forced eccentricity of the odd-degree tesseral terms, the mechanism of the translation theorem. Rederived in Kaula form, its amplitude is, to leading order, independent of $e$ and $\omega$, is amplified by $n/(m\dot\theta)$, and equals about 25~km of perilune; the closed-form degree-3 amplitudes agree with numerical averaging at fixed elements within 2\% and with single-averaged propagation within 5\%. In the force-model decomposition and the term-group runs, $C_{22}$, the Earth and the Sun do not erode the margin.
\item The node dependence of the lowest perilune is the difference between osculating and mean initial eccentricity. A nonsingular single-averaged model reproduces the mission-minimum perilune altitude of all twenty runs within 0.5~km (rms 0.27~km), and a two-step prediction without fitted parameters within 1.2~km for horizons up to 700 days, after which it overestimates the lowest perilune. Targeting the mean eccentricity at insertion reduces the spread over the node to 0.7~km in the first 300 days and raises the three-year worst case from 31.6~km to 39.9~km; with a 70~km mean perilune the lowest altitude stays above 57~km.
\item The slow eccentricity vector rotates about a frozen center whose eccentricity converges only near degree~50 of the zonal series (0.024 at $80.5^\circ$; 0.018 with the Earth). The nominal eccentricity is close to twice the value with the Earth, so the mean vector passes close to the origin, the orbit is close to circular for part of each month in the middle of the mission, and only 6 of the 20 runs belong to the circulating group (5 of the 12 node-scan runs meet R4 under the nominal 27.3-day filter over days 14--1049, and 2--4 over a shorter common span, depending on the filter). A $20\times20$ field misplaces the center and overstates both the circulation and the coverage (78--96\%). The nominal eccentricity lies close to the circular-contour estimate $2e_c$ (0.0360--0.0369, depending on the node) and within the node-dependent range of the origin-crossing threshold at $\omega_0=90^\circ$ of the fixed-$i,\Omega$ reduced model (0.0374--0.0383; 0.0377 in an independent doubly averaged calculation), so the circulation is marginal and depends on the node; since these thresholds refer to the mean eccentricity, they indicate the proximity to the transition but do not classify individual runs; a smallest slow eccentricity along the trajectory larger than the forced amplitude, $\bar e_{\min}>|\delta\varepsilon|_{\max}$, keeps the orbit clear of near-circular passages.
\item The results reduce to three checks that need only the single-averaged model: the circulation of the slow phase, the margin against near-circular passages, and a perilune screen for targeted orbits, which was below the observed lowest perilune in all eight targeted GMAT runs. In the month-averaged model the frozen center approaches the origin closely near $76.3^\circ$ and $84.5^\circ$ inclination, where its signed coordinate changes sign, so the inclination, and not only the eccentricity, is expected to control the circulation; this prediction has not been tested here with full-force propagation.
\item For this orbit the plane precesses about an effective axis inclined about $0.75^\circ$ to the lunar spin axis, because the spin axis itself precesses with an angular rate only three times smaller than the nodal rate of the orbit. Initial states should be defined with the rotation model of the propagator, and prediction models must rotate the field exactly as the propagator does; an axis error of $0.6^\circ$ changed the predicted perilunes by up to 5~km.
\end{enumerate}
A companion paper turns these checks into a design method for a prescribed altitude shell: it gives the frozen center, the shape of the slow contour and its rotation rate in closed form, derives from them propagation-free bounds on the lowest altitude and on the coverage, and verifies the resulting designs with full-force propagation.

\section*{CRediT authorship contribution statement}
\textbf{Fu-Yuen Hsiao:} Conceptualization, Methodology, Formal analysis, Writing -- original draft, Writing -- review \& editing, Supervision, Project administration, Funding acquisition.
\textbf{Kaiti Wang:} Conceptualization (science requirements), Investigation (whistler-wave analysis), Writing -- review \& editing.
\textbf{Wei-Cheng Chou:} Software, Validation, Investigation (including the literature survey), Writing -- review \& editing.

\section*{Declaration of competing interest}
The authors declare that they have no known competing financial interests or personal relationships that could have appeared to influence the work reported in this paper.

\section*{Declaration of generative AI and AI-assisted technologies in the\\ manuscript preparation process}
During the preparation of this work the authors used the following tools. Claude (Anthropic) was used to draft and edit the text, to develop and check analytical derivations, to write and run the analysis code (single-averaged model, prediction and figure scripts) and to prepare the GMAT input scripts; its use in the research itself is also stated in Section~\ref{sec:savg}. Gemini (Google) and Perplexity (Perplexity AI) were used to review the draft manuscript, including the literature positioning and the references. The derivations were checked by the authors, the numerical results were verified against independent full-force GMAT propagations, and the bibliographic data of the references with a DOI were checked against the Crossref records. After using these tools, the authors reviewed and edited the content as needed and take full responsibility for the content of the published article.

\section*{Acknowledgments}
This work was supported by the Taiwan Space Agency (TASA) under Project TASA-S-1140041. The authors thank Sheng-Hung Chen, You-Jun Jian, Yu-Chuan Lai and Chung-Yi Liao (listed in alphabetical order of surname), undergraduate students of the Department of Aerospace Engineering, Tamkang University, for carrying out the GMAT simulations, and Dr.\ Hao-Chi Chang of TASA for reviewing the description of TASA's deep-space exploration plans in this paper.

\appendix
\section{Initial states of the GMAT runs}
\label{app:states}
Table~\ref{tab:app_states} lists the initial Cartesian states of the node-scan and targeting runs at 2043-09-10 00:00:00 UTC in the GMAT \texttt{luna} frame (origin at the Moon, \texttt{BodyInertial} axes), in km and km/s. The values are rounded to 0.1~m and 0.1~mm/s; the runs use the full-precision values of the GMAT scripts.

\begin{table}[htbp]
\centering
\footnotesize
\setlength{\tabcolsep}{4pt}
\caption{Initial Cartesian states in the GMAT \texttt{luna} frame at 2043-09-10 00:00:00 UTC (rounded).}
\label{tab:app_states}
\begin{tabular}{lrrrrrr}
\toprule
Run & $X$ (km) & $Y$ (km) & $Z$ (km) & $V_X$ (km/s) & $V_Y$ (km/s) & $V_Z$ (km/s) \\
\midrule
T000 & -103.4869 & -164.8157 & 1787.4368 & 1.4244431 & -0.8944002 & 0.0000000 \\
T030 & 43.2803 & -210.1019 & 1785.1577 & 1.6800687 & -0.0635229 & -0.0482087 \\
T060 & 192.9350 & -176.0233 & 1778.9311 & 1.4855211 & 0.7843753 & -0.0834999 \\
T090 & 305.3773 & -71.7112 & 1770.4254 & 0.8929294 & 1.4221008 & -0.0964174 \\
T120 & 350.4784 & 74.8842 & 1761.9198 & 0.0610780 & 1.6787755 & -0.0834999 \\
T150 & 316.1536 & 224.4825 & 1755.6932 & -0.7871392 & 1.4856237 & -0.0482087 \\
T180 & 211.6000 & 336.9992 & 1753.4141 & -1.4244431 & 0.8944002 & 0.0000000 \\
T210 & 64.8328 & 382.2855 & 1755.6932 & -1.6800687 & 0.0635229 & 0.0482087 \\
T240 & -84.8219 & 348.2069 & 1761.9198 & -1.4855211 & -0.7843753 & 0.0834999 \\
T270 & -197.2642 & 243.8947 & 1770.4254 & -0.8929294 & -1.4221008 & 0.0964174 \\
T300 & -242.3653 & 97.2994 & 1778.9311 & -0.0610780 & -1.6787755 & 0.0834999 \\
T330 & -208.0404 & -52.2990 & 1785.1577 & 0.7871392 & -1.4856237 & 0.0482087 \\
M000 & -185.8762 & -112.9685 & 1786.5397 & 1.4158066 & -0.9005937 & 0.0903568 \\
M060 & 676.7243 & 92.5449 & 1667.2331 & 1.3538567 & 0.7952166 & -0.5936668 \\
M120 & 355.3803 & 466.7077 & 1694.9401 & -0.0125766 & 1.6260827 & -0.4451106 \\
M180 & 273.7018 & 296.4020 & 1747.4286 & -1.4185327 & 0.9089149 & 0.0680146 \\
M240 & 40.7528 & 411.3633 & 1740.7829 & -1.4941860 & -0.7597568 & 0.2145175 \\
M300 & -204.3144 & 724.0351 & 1637.1456 & -0.1366755 & -1.5363188 & 0.6623870 \\
H060 & 706.9398 & 108.6834 & 1664.3478 & 1.3367733 & 0.7903427 & -0.6194110 \\
H300 & -202.2187 & 763.1598 & 1630.3234 & -0.1401544 & -1.5155803 & 0.6920640 \\
\bottomrule
\end{tabular}
\end{table}

\bibliographystyle{elsarticle-num}
\bibliography{my_bib}

\end{document}